\documentclass[aps,prd,amsmath,amssymb,10pt,letterpaper,balancelastpage,showpacs,superscriptaddress,nofootinbib,floatfix,notitlepage,longbibliography,twocolumn
]{revtex4-2}
\usepackage[utf8]{inputenc}
\usepackage{bm}

\usepackage[dvipsnames]{xcolor}
\definecolor{red}{rgb}{0.9, 0,0}
\definecolor{cerulean}{rgb}{0., 0.42,0.9}
\definecolor{navy}{rgb}{0.05, 0.05,0.8}

\usepackage[colorlinks]{hyperref}
\hypersetup{
  colorlinks = true,
  citecolor = red,
	linkcolor = navy
}
\usepackage{comment}
\usepackage{caption}
\usepackage{subcaption}

\usepackage{bbm}
\usepackage{amsfonts}
\usepackage{mathrsfs}
\usepackage{bbold}
\usepackage{amsmath,amssymb}
\usepackage{siunitx}
\usepackage{slashed}
\usepackage{epsfig}
\usepackage{graphicx}   
\usepackage{url}
\usepackage{float}
\usepackage{soul}
\usepackage{pstricks}
\usepackage{color}
\usepackage{multirow}
\usepackage{mathrsfs} 
\usepackage[normalem]{ulem} 
\usepackage{enumitem}
\usepackage{multirow}
\usepackage{diagbox}

\newcommand{\eq}[1]{Eq.~(#1)}
\newcommand{\Eq}[1]{Eq.~(#1)}

\usepackage[
    starfontserif 
    ]{starfont}
\DeclareSymbolFont{starfontsym}{OT1}{sts}{m}{n}
\DeclareMathSymbol{\mathSun}{\mathord}{starfontsym}{115}
\DeclareMathSymbol{\mathMercury}{\mathord}{starfontsym}{102}
\DeclareMathSymbol{\mathVenus}{\mathord}{starfontsym}{103}
\DeclareMathSymbol{\mathTerra}{\mathord}{starfontsym}{76}
\DeclareMathSymbol{\mathvarTerra}{\mathord}{starfontsym}{108}
\DeclareMathSymbol{\mathMoon}{\mathord}{starfontsym}{100}
\DeclareMathSymbol{\mathvarMoon}{\mathord}{starfontsym}{97}
\DeclareMathSymbol{\mathMars}{\mathord}{starfontsym}{104}
\DeclareMathSymbol{\mathJupiter}{\mathord}{starfontsym}{106}
\DeclareMathSymbol{\mathSaturn}{\mathord}{starfontsym}{83}
\DeclareMathSymbol{\mathUranus}{\mathord}{starfontsym}{70}
\DeclareMathSymbol{\mathvarUranus}{\mathord}{starfontsym}{65}
\DeclareMathSymbol{\mathNeptune}{\mathord}{starfontsym}{71}
\DeclareMathSymbol{\mathPluto}{\mathord}{starfontsym}{74}
\DeclareMathSymbol{\mathvarPluto}{\mathord}{starfontsym}{72}

\newcommand{\vbold}{\mathbf{v}}

\newcommand{\rmax}{r_{\text{max}}}

\newcommand{\Vcen}{V_{\text{cen}}}
\newcommand{\Ncen}{N_{\text{cen}}}
\newcommand{\Vsam}{V_{\text{sample}}}
\newcommand{\gt}{\tilde{g}}
\newcommand{\Lint}{L_{\text{int}}}
\newcommand{\mg}{m_{\gt}}

\newcommand{\TeV}{\text{TeV}}
\newcommand{\teq}{t_{\text{eq}}}

\newcommand{\Ncycle}{N_{\text{cycle}}}
\newcommand{\Vfinal}{V_{\text{final}}}

\newcommand{\vth}{v_{\text{th}}}
\newcommand{\vdrift}{v_{\text{drift}}}
\newcommand{\lambdatr}{\lambda_{\text{tr}}}
\newcommand{\tautr}{\tau_{\text{tr}}}
\newcommand{\sigmatr}{\sigma_{\text{tr}}}
\newcommand{\trmax}{\tilde{r}_{\text{max}}}
\newcommand{\tl}{\tilde{l}}
\newcommand{\vlab}{v_{\text{lab}}}
\newcommand{\vcm}{v_{\text{CM}}}
\newcommand{\vstop}{v_{\text{stopped}}}
\newcommand{\vionmax}{v_{\text{ionmax}}}
\newcommand{\lambdaint}{\lambda_{\text{int}}}
\newcommand{\xstop}{x_{\text{stopped}}}
\newcommand{\xthr}{x_{\text{thr}}}
\newcommand{\vesc}{v_{\text{esc}}}
\newcommand{\sentry}{s_{\text{entry}}}
\newcommand{\sexit}{s_{\text{exit}}}
\newcommand{\scylinder}{s_{\text{cylinder}}}
\newcommand{\rhoatlas}{\rho_{\text{ATLAS}}}
\newcommand{\Nsample}{N_{\text{sample}}}
\newcommand{\cmin}{c_{\text{min}}}
\newcommand{\Estop}{
\mathcal{E}_{\text{stopped}}}

\begin{document}

\title{Melting LHC detectors: A novel search for stopped long-lived particles}

\author{Julia L.~Gonski}
\affiliation{SLAC National Accelerator Laboratory, Menlo Park, California 94025, USA}
    
\author{Peter W.~Graham}
\affiliation{Leinweber Institute for Theoretical Physics, Department of Physics, Stanford University, Stanford, California 94305, USA}
\affiliation{Kavli Institute for Particle Astrophysics and Cosmology, Department of Physics, Stanford University, Stanford, California 94305, USA}

\author{Surjeet Rajendran}
\affiliation{William H. Miller III Department of Physics and Astronomy,
Johns Hopkins University, Baltimore, Maryland 21218, USA}

\author{Harikrishnan Ramani}
\affiliation{Department of Physics and Astronomy, University of Delaware and the Bartol Research Institute, Newark, Delaware 19716, USA}

\author{Samuel S.~Y.~Wong}
\email{samswong@uw.edu}
\affiliation{Leinweber Institute for Theoretical Physics, Department of Physics, Stanford University, Stanford, California 94305, USA}
\affiliation{Department of Physics, University of Washington, Seattle, Washington 98195, USA}

\begin{abstract}
Particles at the TeV scale with lifetimes of a year or longer could have been abundantly produced at the LHC yet escaped detection because of backgrounds, and could still be trapped within detector materials. With gluinos in split supersymmetry as a working example, we show that these trapped particles can be recovered from detector materials once prepared in liquid form, for example, by melting silicon detectors, extracting liquid argon from the electromagnetic calorimeter,  or constructing a large water pool near ATLAS or CMS. These liquid samples can then be processed using iterative centrifugation followed by mass spectrometry, enabling single-particle sensitivity in macroscopic samples. This method can potentially discover gluinos up to 3 TeV in mass at the HL-LHC. It can also improve upon existing limits for other long-lived particles. For example, it can discover the stop up to 2 TeV, and will also be sensitive to integer-charged particles in the TeV range.
\end{abstract}

\maketitle

\tableofcontents

\section{Introduction}
\label{sec:introduction}
Long-lived particles (LLPs) arise naturally in many theories beyond the standard model (SM). Their long lifetimes could result from large mediator masses, small couplings to their decay products,  suppressed phase space due to nearly degenerate mass spectra, or exact symmetries that render them stable. Various models predict the existence of LLPs at the TeV scale, and the exciting prospect of producing them at the Large Hadron Collider (LHC) has driven substantial work to extend the LHC’s sensitivity to macroscopic decay length, $\tau \gg 1~\mu \text{m} \sim 1~\text{fs}$~\cite{LLP_Experiments_Survey,Moedal_2009,Mathusla,milliQan_2014,codexb_2017,faser_2017}.

LLPs can be classified by whether they decay on collider timescales. Particles with lifetimes $\tau < 1$~ns may produce displaced vertices, prompting dedicated searches by ATLAS, CMS, and LHCb~\cite{ATLAS_displaced_hadronic_2022,ATLAS_displaced_lepton_2022,ATLAS_displaced_muon_2020,ATLAS_displaced_inner_muon_2019,ATLAS_displaced_hadronic_calorimeter_2022,ATLAS_displaced_dark_photon_2022,ATLAS_displaced_missing_momentum_2017,CMS_displaced_lepton_2022,CMS_displaced_jets_2018,CMS_displaced_muon_2022,LHCb_displaced_2022}. The proposed MATHUSLA detector could extend lifetime sensitivity to $\tau \sim 0.1~\text{s}$~\cite{Mathusla,MATHUSLA_2025}. For much longer lifetimes, $\tau \gg 0.1~\text{s}$, decay-based signatures become increasingly suppressed, vanishing altogether for stable particles. In such cases, if these stable (or quasistable) particles are charged or hadronize into charged states, they are referred to as heavy stable charged particles (HSCPs).

Such HSCPs could be identified through their anomalously large ionization energy losses ($dE/dx$) in the silicon trackers in ATLAS and CMS~\cite{ATLAS_dEdx_2025, ATLAS_dEdx_2022,CMS_dEdx_2024}. This method relies on the fact that HSCPs are typically produced with lower velocities than standard model particles and consequently have higher $dE/dx$, according to the Bethe-Bloch equation. By reconstructing $dE/dx$ profiles from the measured charge deposition along tracks with high transverse momentum in the silicon pixel detectors, and cross-checking with other time-of-flight observables, the experiments could identify an excess of events consistent with slow, heavy particles beyond the expected standard model background~\cite{ATLAS_LLP_Review_Gonski}.

Despite setting the most stringent constraints on much of HSCPs parameter space, the $dE/dx$ method is limited by backgrounds due to pile-up. Particles originating from simultaneous proton-proton (pp) collisions may produce overlapping detector hits, which degrade track reconstruction and increase the rate of false-positive tracks~\cite{ATLAS_track_reconstruction_pileup_2023}. In ATLAS Run 2, the mean pile-up, defined as the number of simultaneous pp collisions per bunch crossing, was $\langle \mu \rangle = 34$~\cite{ATLAS_dEdx_2022,ATLAS_dEdx_2025}.

Furthermore, this problem may be exacerbated at the High-Luminosity LHC (HL-LHC), which aims to achieve an integrated luminosity of $\Lint = 3000~\text{fb}^{-1}$, roughly 20--30 times the Run 2 value. This higher luminosity will proportionally enhance the expected number of signal events; at the same time, the pile-up is expected to rise to $\langle \mu \rangle = 200$~\cite{HLLHC_ATLAS_detector_2019,HLLHC_Yellow_Report_2019,HLLHC_ATLAS_pileup_2013}. It is currently unknown how much the $dE/dx$ search can improve with increased pile-up. As the HL-LHC era approaches, it is important to develop new strategies that maximize the discovery potential afforded by the higher luminosity while mitigating the accompanying rise in background noise~\cite{Bhattacherjee:2025pxg}.

In this context, we introduce a new class of background-free signals of HSCPs that can fully take advantage of the luminosity reach of the HL-LHC: the recovery of trapped particles in detector material. As HSCPs traverse detector components, they lose kinetic energy through ionization and eventually come to rest, provided their initial velocities are sufficiently low. Their electric or color charges then cause them to bind to atomic nuclei or electrons, forming anomalously heavy atoms. Searches for decays of such stopped particles have been carried out in ATLAS~\cite{stopping_gluino,stopped_gluino_decay_ATLAS_2021}, but, again, these strategies are not sensitive to lifetimes much longer than a year. For this regime, an irreducible signal of interest here is the presence of a tiny concentration of such heavy atoms embedded within the detector material. Since all normal atoms are much lighter than the TeV scale, this enables a background-free search.

We propose a precision measurement experiment to search for such heavy atoms. Our method requires the detector material to be first converted into liquid form after the full run of the collider. The liquid sample will then be placed in a high-speed centrifuge, which separates components of the liquid by density through centrifugation. Iterative application of centrifugation, retaining only the heaviest portion at each step, rapidly reduces the liquid volume. The resulting highly enriched, microliter-scale sample can then be analyzed using mass spectrometry to identify trace concentrations of heavy particles. With state-of-the-art instrumentation, this procedure is, in principle, sensitive to a single anomalously heavy atom in the original sample.

Although the ideas introduced here apply to any region surrounding an interaction point, such as ATLAS, CMS, or even future hadron colliders, here we focus on ATLAS and consider three target materials within or near it. First, the silicon detectors~\cite{ATLAS_ID_Run2, ITk_Details_ATLAS_2024}, located closest to the collision point and subject to the highest exposure, are periodically replaced due to radiation damage~\cite{Atlas_ITk_replacement}. These retired detectors could be melted and chemically converted into a liquid before undergoing our analysis. Second, the electromagnetic (EM) calorimeter~\cite{ATLAS_Liquid_Argon} contains a large volume of liquid argon, conveniently already in liquid phase, which could be extracted during detector upgrades or at the end of the detector’s operational lifetime. For our purposes, these are passive detectors, requiring no active monitoring instrumentation, much like the MoEDAL experiment~\cite{Moedal_2009}, and are already collecting possible HSCPs during LHC operations, without requiring new infrastructure. 

We also consider a more ambitious target: a hypothetical, custom-built volume of water placed outside ATLAS, which has the advantage of a much larger target volume. While it requires some investment in infrastructure, the cost is almost entirely limited to excavation, with no ongoing maintenance required.

Similar ``heavy isotope" searches in terrestrial materials were conducted in the past~\cite{heavy_hydrogen_water_1967,heavy_hydrogen_1977,heavy_hydrogen_1978,heavy_oxygen_1979,HeavyWater1,HeavyWater2,heavy_isotope_1981,heavy_carbon_1984,heavy_sodium_1984,heavy_iron_1987,strange_matter_Rutherford_1989,sea_water_centrifuge_proposal_1987,sea_water_centrifuge_1992,heavy_isotopes_lowZ_1990,heavy_nuclei_search_Rutherford_1991,deep_sea_water_1993,simp_search_2001} and reviewed in Refs.~\cite{stable_search_review_2001,strangelets_search_review_2005,noncollider_searches_review_2014}. Those studies targeted, among others, HSCPs hypothesized to originate as early universe relics, which persist in small abundances within ordinary terrestrial matter such as seawater. This work pursues a variation of that research program, with the production mechanism instead being the pp collisions at the LHC, which have lower energy reach than cosmological processes but whose occurrence is certain and localized.\footnote{In the review Ref.~\cite{noncollider_searches_review_2014}, it was mentioned that ``Somewhat surprisingly, no searches have been made for electrically charged [stable massive particles] trapped in
matter at high-energy colliders."} Notably, the use of centrifuges to enrich seawater as a means to search for heavy particles was performed in Refs.~\cite{sea_water_centrifuge_proposal_1987,sea_water_centrifuge_1992}, and the time-of-flight mass spectrometry procedure we outline here is based on the experiment conducted in Refs.~\cite{HeavyWater1,HeavyWater2}.

We study the potential physics reach of our method by focusing on the long-lived gluino in split supersymmetry as our benchmark particle, reviewed in Sec.~\ref{sec: gluino}. Section~\ref{sec: tapping gluinos} details our simulation of gluino production and its subsequent stopping in the LHC. The proposed search for heavy particles in liquid samples is described in Sec.~\ref{sec:heavysearch}, and the projected reach for gluino detection is presented in Sec.~\ref{sec:projection}.

\section{Long-lived gluino}
\label{sec: gluino}
In this paper, we focus on the long-lived gluino in split supersymmetry (SUSY)~\cite{split_susy,split_susy_conditions,split_susy_little_hierarchy_2003} as a well-motivated HSCP benchmark, though our experimental technique applies more generally. In split SUSY, the Higgs mass is fine-tuned, and the SUSY-breaking scale is essentially a free parameter assumed to be far above the weak scale, $m_S \gg 1~\text{TeV}$, driving up the masses of all scalar sparticles. The gauginos and Higgsinos remain at the TeV scale, motivated by gauge coupling unification~\cite{unification_susy_su5_1981,unification_susy_gut_1981,Marciano:1981un,Einhorn:1981sx,Ibanez:1981yh,Sakai:1981gr,Langacker:1995fk} and the identification of the lightest neutralino, stabilized by R-parity, as a candidate for the weakly interacting massive particle (WIMP) dark matter~\cite{LSP_1984,susy_dm_review, LastWIMP,EnhancingHiggsinoDM_2024,Higgsino_CTAO_2025}.

In this model, the gluino ($\gt$) is the only new TeV-scale colored particle and hence the only sparticle copiously produced at the LHC. A distinctive feature of split SUSY is the longevity of the gluino~\cite{split_susy}. Due to R-parity, it must decay through a virtual squark exchange~\cite{gluino_decay_diagrams_2005}, and its lifetime is thus suppressed by the large squark masses at scale $m_S$,
\begin{align}
    \tau_{\gt} = 0.4 ~\text{year} \left(\frac{1}{N}\right) 
    \left(\frac{m_S}{10^{11} ~\text{GeV}} \right)^4
    \left(\frac{2~\text{TeV}}{\mg} \right)^5  ~,
\end{align}
where $N$ is an order-one function~\cite{gluino_decay_split_2005}.
Due to the steep scaling in this formula, the gluino lifetime could easily range from collider to cosmological timescales.

The current limit on stable or long-lived gluino mass, $\mg > 2.20$~TeV, is set by the ATLAS $dE/dx$ search~\cite{ATLAS_dEdx_2025}. However, gluino masses as high as 3~TeV would be produced in significant numbers ($\sim 60$ pairs at the HL-LHC). We thus focus on the mass range $\mg =2$--3~TeV and lifetimes longer than roughly a year, corresponding to probing a SUSY-breaking scale of approximately $m_S > 10^{11}$~GeV. In the model with minimal field contents~\cite{mini_split_2012,unnatural_susy_2012}, the observed Higgs mass favors a smaller $m_S$, but this need not be the case for more generic models~\cite{split_susy}.

For $\mg = 2$--3~TeV explored here, gluino lifetimes shorter than $3\times 10^4$ years or longer than $10^{15}$ years are consistent\footnote{An earlier treatment~\cite{gluino_cosmology_bounds_2005} neglected nonperturbative postconfinement annihilation and therefore overestimated the gluino relic density.}\textsuperscript{,}\footnote{TeV-scale leptons with lifetimes in the 10--100 year range are constrained by the CMB~\cite{4th_gen_Sher_1992,4th_gen_Sher_2000}; a similar bound might apply to the gluino.} with cosmological constraints~\cite{gluino_cosmology_bounds_2018}.  Note that if the reheating temperature was below $\sim \TeV$~\cite{reheat_lowest_2004}, gluinos would not have been produced, and these cosmological limits would not apply at all~\cite{split_susy}. 

The gluino lifetime can be longer than the age of the universe because the null results of the heavy isotope searches~\cite{heavy_hydrogen_water_1967,heavy_hydrogen_1977,heavy_hydrogen_1978,heavy_oxygen_1979,HeavyWater1,HeavyWater2,heavy_isotope_1981,heavy_carbon_1984,heavy_sodium_1984,heavy_iron_1987,strange_matter_Rutherford_1989,sea_water_centrifuge_proposal_1987,sea_water_centrifuge_1992,heavy_isotopes_lowZ_1990,heavy_nuclei_search_Rutherford_1991,deep_sea_water_1993,simp_search_2001} cannot be interpreted as limits on gluino lifetime, as also pointed out in Ref.~\cite{colored_dm_2018}. While these searches offered promising discovery potential, setting an actual bound on the gluino lifetime would require a detailed model of how gluinos are mixed with normal matter and transported throughout cosmological, galactic, Solar System, and planetary evolution to end up in terrestrial matter. Any claim of a limit is therefore premature until
such modeling becomes possible; even then, it remains subject to the reheating temperature caveat. These considerations reveal a viable region of gluino parameter space that can be probed by the method introduced in this work.

\section{Trapping gluinos at the LHC}
\label{sec: tapping gluinos}
\subsection{Gluino production}
We simulated gluino pair production, $pp\!\to\!\tilde{g} \tilde{g}$, at the LHC with a center-of-mass energy of $\sqrt{s} = 13.6$~TeV~\cite{ATLAS_run3}, at tree level. The simulation used the MadGraph5\_aMC event generator with the MSSM\_SLHA2 model in the decoupled-squark limit~\cite{MG5aMC,SUSY_Les_Houches_Accord_2}. 
For each gluino mass in the range $\mg=$~2--3~TeV, varied in steps of 100~GeV, we generated $5 \times 10^5$ events, corresponding to $10^6$ produced gluinos. The simulation provides the shapes of the gluino momentum distributions but we take the total production cross section from Refs.~\cite{gluino_cross_section_13.6,gluino_cross_section_13}. 

Gluinos with low lab-frame velocity, $\vlab \lesssim 0.1$, are of particular interest because they can stop in detector material.  Because $\vlab$ correlates with the center-of-mass (CM) velocity $\vcm$, our result is sensitive to the production cross section for gluino pairs with low $\vcm$. This cross section is enhanced by resumming Sommerfeld ladder diagrams arising from  virtual gluons exchange in the final state~\cite{Sommerfeld1931,AkhiezerBerestetskii1965,Schwinger1973}.
The Sommerfeld enhancement factor is
\begin{align}
    E_S(\vcm) = \frac{C\pi \alpha_s}{\vcm} \left[ 1- \exp\left(- \frac{C \pi \alpha_s}{\vcm} \right) \right]^{-1} ~,
\end{align}
where $\vcm = \sqrt{1-4\mg^2/\hat{s}}$~\cite{Sommerfeld_Factor} with $\hat{s}$ the partonic Mandelstam variable. We provide further details on this formula and its numerical treatment in Appendix~\ref{sec: Sommerfeld}. 

Soft gluon emission also generates a threshold enhancement, a low velocity effect that scales as $\alpha_s \log^2 (\vcm^2)$ at next-to-leading order (NLO)~\cite{threshold_resum_estimate_2009}. Resumming these logarithms is more involved and is usually not carried out in position space~\cite{threshold_coulomb_resum_2016}, so we conservatively omit them here.\footnote{However, the total cross section considered here does include threshold resummation~\cite{gluino_cross_section_13.6,gluino_cross_section_13}.} However, we checked that the NLO correction incorporating both soft gluon emission and virtual gluon exchange is subdominant to the Sommerfeld resummation considered here~\cite{threshold_coulomb_NLO_1996}.

\begin{figure}[t]
\centering
\begin{minipage}{0.492\textwidth}
  \centering
  \includegraphics[width=\linewidth]{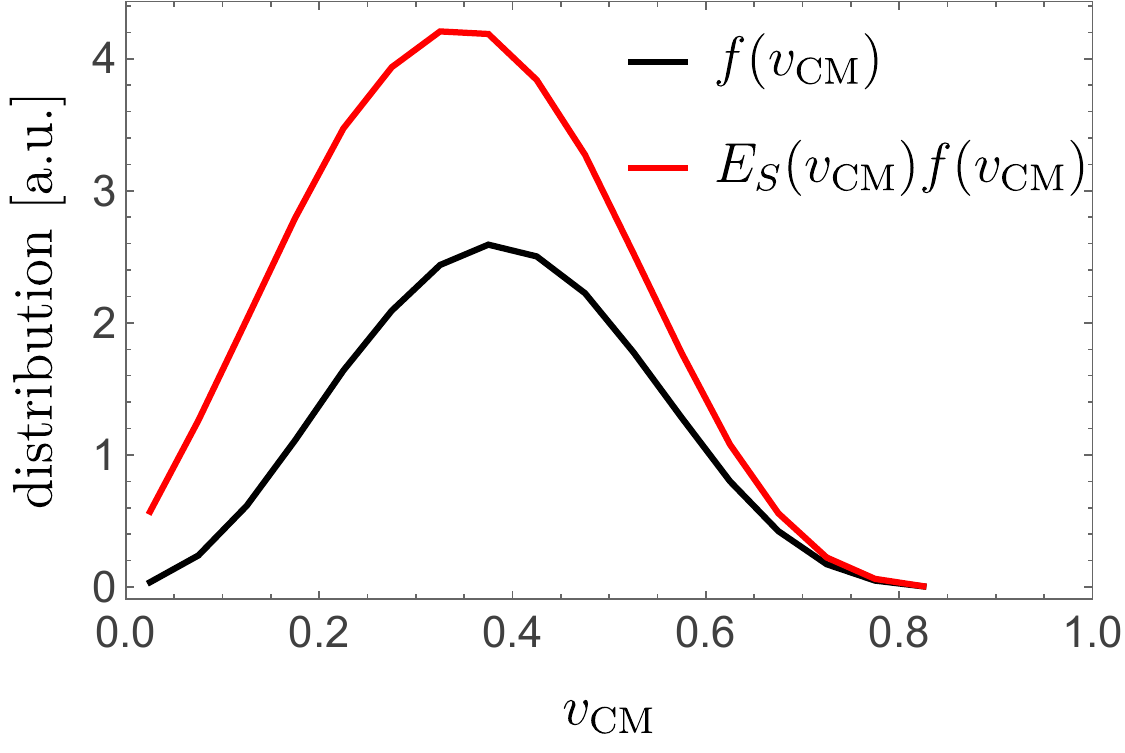}
  \subcaption{}
\end{minipage}
\begin{minipage}{0.492\textwidth}
  \centering
    \includegraphics[width=\linewidth]{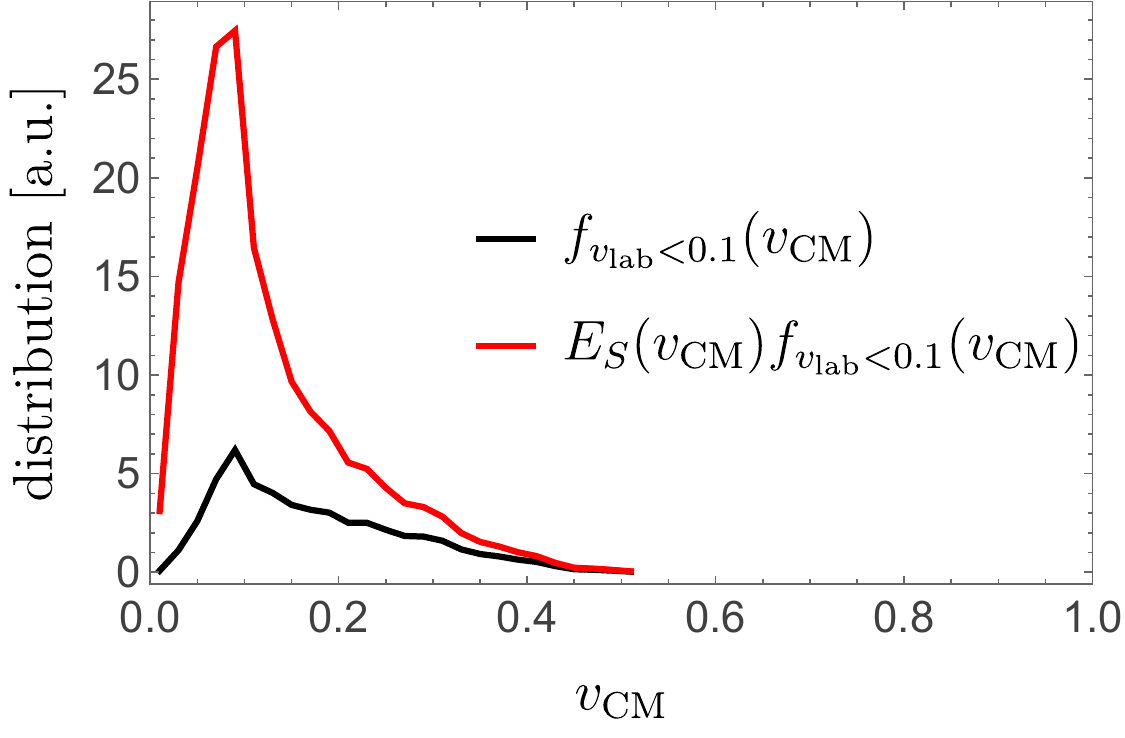}
    \subcaption{}
\end{minipage}
\caption{(a) Velocity distribution $f(\vcm)$ of a 3 TeV gluino in the CM frame and its Sommerfeld-enhanced distribution $E_S(\vcm) f(\vcm)$, in arbitrary units (a.u.). (b) The distribution for the subset of gluinos whose lab-frame velocity satisfies $\vlab<0.1$, roughly the population stopping within the detector.}
\label{fig: Sommerfeld distribution}
\end{figure}

\newcommand{\figSize}{0.45}
\begin{figure}[t]
\centering
\begin{minipage}{\figSize \textwidth}
  \centering
  \includegraphics[width=\linewidth]{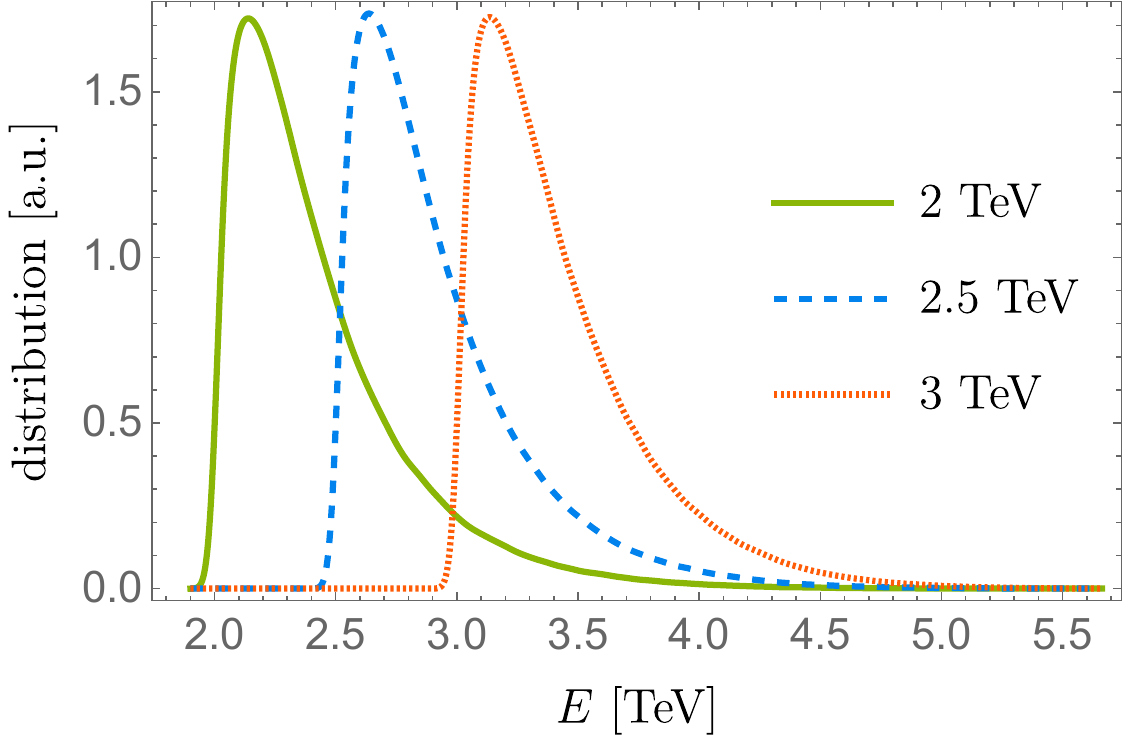}
  \subcaption{}
  \label{fig: E Distribution}
\end{minipage}
\begin{minipage}{\figSize \textwidth}
  \centering
\includegraphics[width=\linewidth]{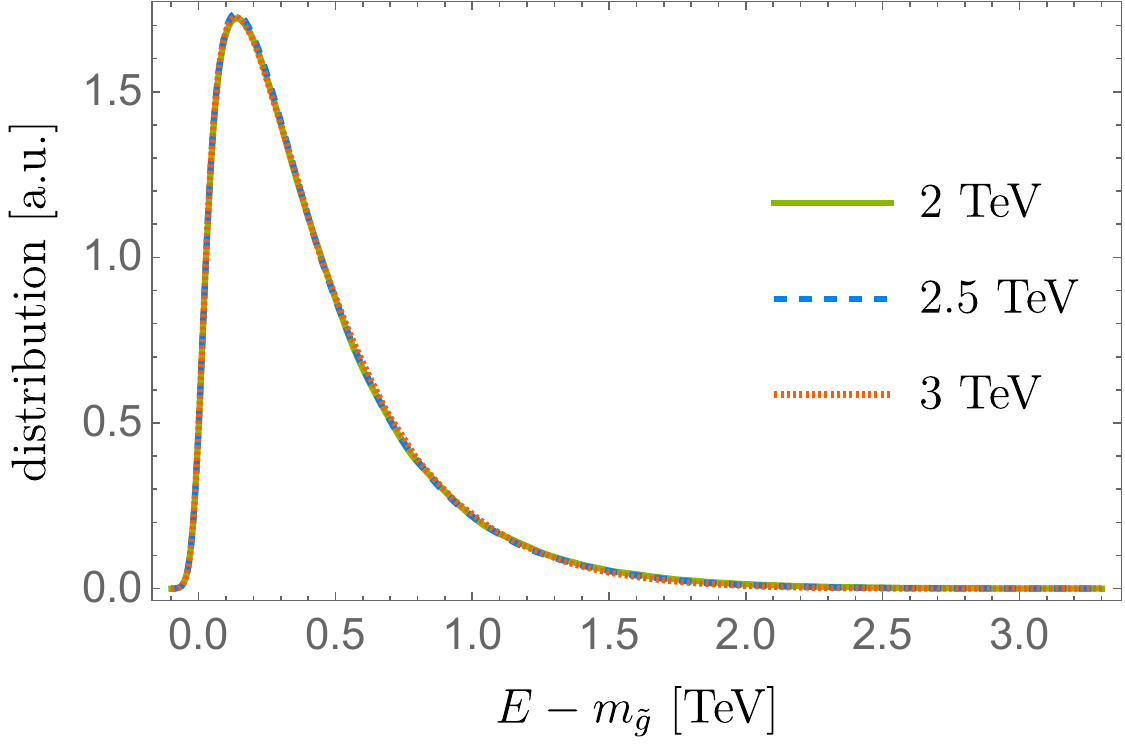}
    \subcaption{}
    \label{fig: E-m Distribution}
\end{minipage}
\begin{minipage}{\figSize \textwidth}
  \centering
\includegraphics[width=\linewidth]{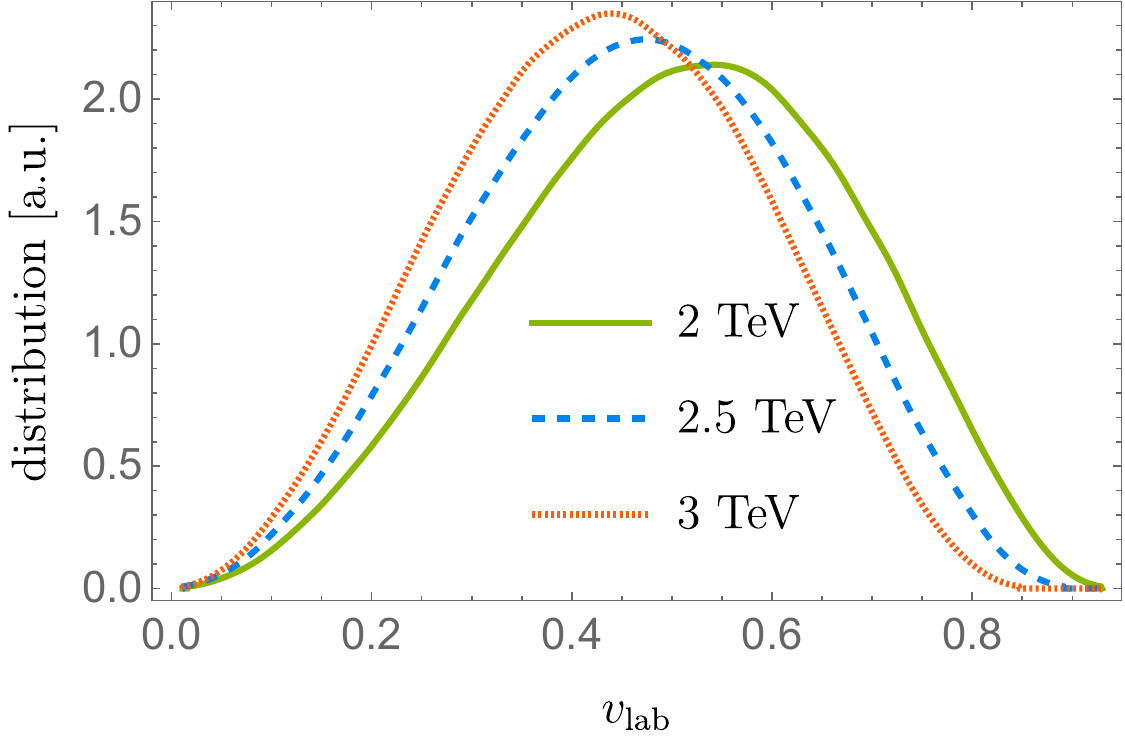}
    \subcaption{}
    \label{fig: vlab Distribution}
\end{minipage}
\caption{(a) The distributions of the lab-frame energy $E$ for various gluino masses $\mg$ have the same shape and only shifted in the energy axis. (b) The kinetic energy distributions $E-\mg$ for the three masses are identical. (c) The distributions of the lab-frame velocity.}
\label{fig: Kinetic Energy Distribution}
\end{figure}

We modified the simulated gluino distributions from MadGraph5 to account for Sommerfeld enhancement. For a population of $N_{\text{sim}}$ simulated gluinos with a normalized velocity distribution $f(\vcm)$, we define the augmented number $\bar{N}_{\text{sim}}$ and the integrated enhancement factor $E_S[f]$ by
\begin{align} \label{eq: Nsim bar}
    \bar{N}_{\text{sim}} &\equiv N_{\text{sim}} \times \underbrace{\int_0^1 d\vcm E_S(\vcm) f(\vcm)}_{\equiv E_S[f]}
    ~.
\end{align}
As shown in Fig.~\ref{fig: Sommerfeld distribution}, the distribution of 3~TeV gluinos is enhanced predominantly in the low velocity region.  With the additional cut $\vlab<0.1$, roughly the population that would stop in the detector, the enhancement increases by a factor $E_S[f_{\vlab<0.1}]/E_S[f] \approx 2$.

The kinetic energy distribution in the $\mg=$~2--3~TeV mass range was found to be independent of the gluino mass, with a peak at $E-\mg \approx 140$~GeV, as shown in Fig.~\ref{fig: Kinetic Energy Distribution}. The peak velocity is about $\vlab \sim 0.4$--0.5, in the semirelativistic regime. Parton showers after production may further slow the gluinos, potentially increasing the stopping fraction and improve our results; we conservatively neglect this effect~\cite{shower_review_2011}.

\subsection{Hadronization and charge oscillation}
Upon production, gluinos bind with gluons or light quarks to form color-singlet hadrons (R-hadrons). These states span a range of isospin multiplets, approximately half of which are electrically charged ($\pm e$) and half neutral~\cite{stopping_gluino}; we conservatively neglect the possibility of a doubly charged R-baryon state ($\gt uuu$).

As R-hadrons traverse detector materials, they undergo nuclear collisions about once every nuclear interaction length $\lambda_{\text{int}}$ (typical values: $\lambda_{\text{int}} = 47$~cm for silicon and $\lambda_{\text{int}}= 18$~cm for lead)~\cite{PDG_material_properties}. These interactions have minimal effect on their momenta but can alter their quark contents, inducing transitions between charged and neutral states. Consequently, an R-hadron spends roughly half of its time in a charged state while propagating through matter. 

An R-hadron loses energy via ionization and atomic excitation while charged~\cite{stopping_gluino,ATLAS_Rhadron_Simulation_2019}. For simplicity, we approximate the stopping power experienced by the R-hadron as half that of a charged particle. Although the mass spectrum of R-hadrons is difficult to calculate from first principles, Ref.~\cite{stopping_gluino} enumerates possible hierarchies and argues that this average charge fraction is largely insensitive to the specific spectrum. We also conservatively neglect the possibility that, before stopping, an R-hadron binds to a nucleus and forms a multiply charged state, which would only increase its stopping power.

\subsection{Electromagnetic energy loss}
\label{sec: EM deceleration}
The electromagnetic energy loss per unit length of a charged particle is given by the Bethe-Bloch equation~\cite{PDG2024_dEdx,stopping_gluino}
\begin{align} \label{eq: dEdx}
\frac{dE}{dx}=-\underbrace{\frac{4\pi \alpha^2 \rho Z z^2}{A m_p m_e}}_{\equiv C} \frac{1}{v^2} \left[ \log\left( \frac{2m_e v^2}{I (1-v^2)}\right) - v^2\right] ~,
\end{align}
where $v \equiv \vlab$ is the lab-frame velocity of the charged particle, $z$ is its electric charge in units of $e$, $\rho$ is the material density, $I$ is the material-dependent mean excitation energy (Table~\ref{tab:stopping_I}), and, assuming a single-element material, $Z$ and $A$ are its atomic and mass number, respectively.\footnote{For a compound material, Bragg additivity dictates that the stopping power is a weighted sum by mass fraction, $\frac{dE}{dx} = \sum_j w_j \left. \frac{dE}{dx} \right|_j$, where $w_j$ is the fractional mass of the $j$-th element of the material~\cite{PDG2024_dEdx}. Other constants in \eq{\ref{eq: dEdx}}: $\alpha$, $m_p$, and $m_e$ denote the fine structure constant and the proton and electron masses, respectively.}

For the gluino,\footnote{In what follows, we use the terms gluinos and R-hadrons interchangeably, wherever the context permits.} we take $z=1$ but multiply $dE/dx$ by a factor of $\xi = 1/2$ to take into account its charge oscillation (Ref.~\cite{stopped_gluino_decay_ATLAS_2021} also assumes $\xi \approx 1/2$). Note this is different from having an effective charge of $z=1/2$ since the stopping power is proportional to $z^2$.

\begin{table}[t]
    \centering
    \renewcommand{\arraystretch}{1.2} 
    \begin{tabular}{c c @{\hspace{1em}} c}
        \hline
        Material & $I$ [eV] & $\kappa$ \\
        \hline
        water              & 75   & 4.76 \\
        carbon             & 78   & 4.74 \\
        aluminum           & 166  & 4.36 \\
        silicon            & 173  & 4.34 \\
        argon              & 188  & 4.30 \\
        iron               & 286  & 4.09 \\
        copper             & 322  & 4.03 \\
        lead               & 823  & 3.56 \\
        \hline
    \end{tabular}
    \caption{Mean excitation energy \( I \) and parameter \( \kappa \) for various materials used in stopping calculations~\cite{NIST_Table}.}
    \label{tab:stopping_I}
\end{table}

We rewrite the Bethe-Bloch equation as
\begin{align}
    \frac{dv}{dx} = \frac{(1-v^2)^{3/2}}{\mg v} \frac{dE}{dx}~.
\end{align}
In the nonrelativistic limit, this becomes~\cite{stopping_gluino}
\begin{align} \label{eq: dvdx}
    \frac{dv}{dx} \approx - \frac{\xi}{x_0 v^3} \left( 1 + \frac{\log v}{\kappa} \right) ~,
\end{align}
where the parameters are defined as
\begin{align}
    \kappa &\equiv \frac{1}{2} \log \left( \frac{2m_e}{I} \right)~, \\
    \frac{1}{x_0} &\equiv \frac{2 C \kappa}{\mg}~,\\
        \xi &\equiv
    \begin{cases}
        \dfrac{1}{2}, & \text{charge oscillation,} \vspace{0.5 em} \\
        1, & \text{no charge oscillation.}
    \end{cases}
\end{align}

Equation~\ref{eq: dvdx} is only applicable for velocities above $v=\vionmax$, where the ionization reaches its maximum and the particle becomes an adiabatic perturbation on the atomic electrons~\cite{stopping_gluino}.
The point at which this breakdown occurs is given by the condition 
\begin{align} \label{eq: max ion condition}
    \left. \frac{d^2v}{dx^2} \right|_{v=\vionmax} =0 ~,
\end{align}
whose solution is given by\footnote{Note that \Eq{\ref{eq: dvdx}} implies $dv/dx$ vanishes at the velocity $e^{-\kappa}<\vionmax$, but this is unphysical because the velocity exceeds the regime of validity of Bethe-Bloch.}
\begin{align}
    \vionmax = e^{- \kappa+\frac{1}{3}} = e^{\frac{1}{3}} \sqrt{\frac{I}{2m_e}} \approx 0.014 \left(\frac{I}{100~\text{eV}}\right)^{1/2}  ~.
\end{align}
In fact, if we take the ionization energy in the hydrogen model, $I \sim m_e \alpha^2/2$, we get $\vionmax \sim \mathcal{O}(\alpha)$~\cite{stopping_gluino}. 

Deceleration beneath this velocity is described by the Fermi-Teller theory, which states that $dv/dx$ is a constant~\cite{Fermi_Teller,stopping_gluino}, 
\begin{align}
\left. \frac{dv}{dx} \right|_{v=\vionmax} =  - \frac{2 \xi C e^{3\kappa - 1}}{3 \mg} ~,
\end{align}
such that $dv/dx$ is continuous, and \Eq{\ref{eq: max ion condition}} guarantees $d^2v/dx^2$ is also continuous.

\begin{figure}[t] 
     \centering
    \includegraphics[width=0.492\textwidth]{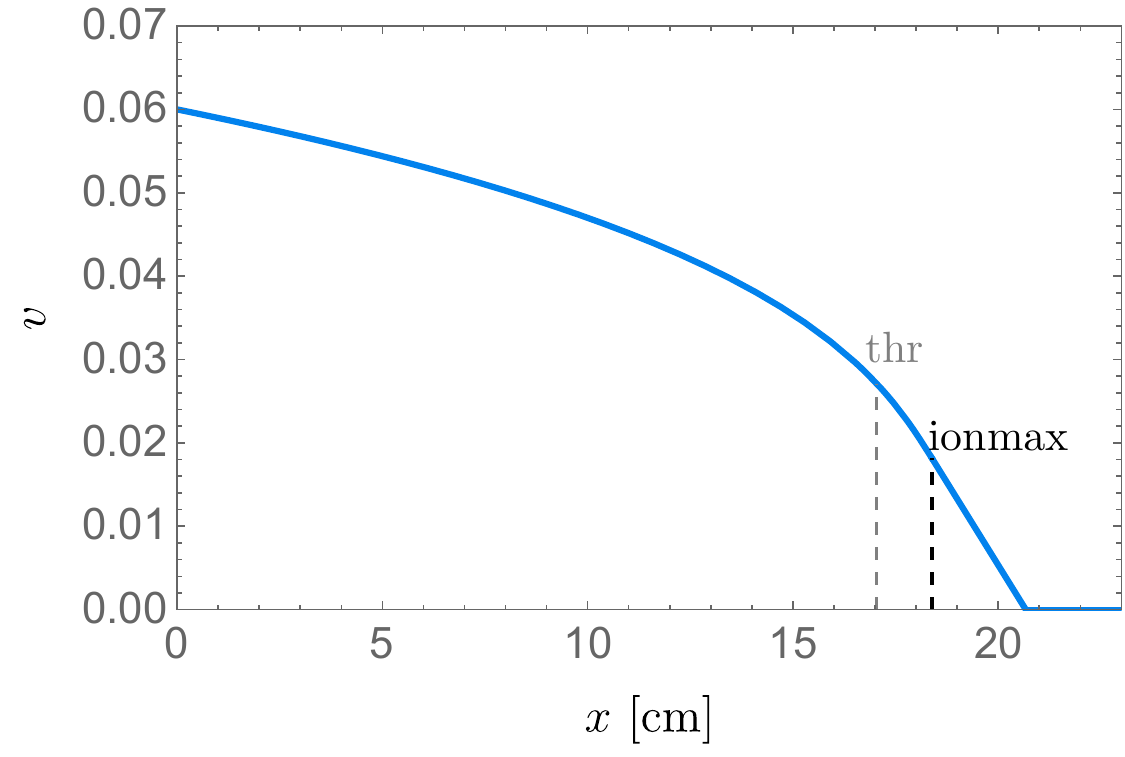}  
 \caption{The velocity $v(x)$ of a 2.5 TeV gluino with initial velocity $v_0=0.06$ traversing through silicon as a function of distance $x$. The point at which ionization reaches its maximum is labeled ``ionmax." The threshold for nuclear stopping is labeled ``thr."}
      \label{fig: gluino deceleration in silicon} 
\end{figure}

\begin{figure}[t] 
     \centering
    \includegraphics[width=0.492\textwidth]{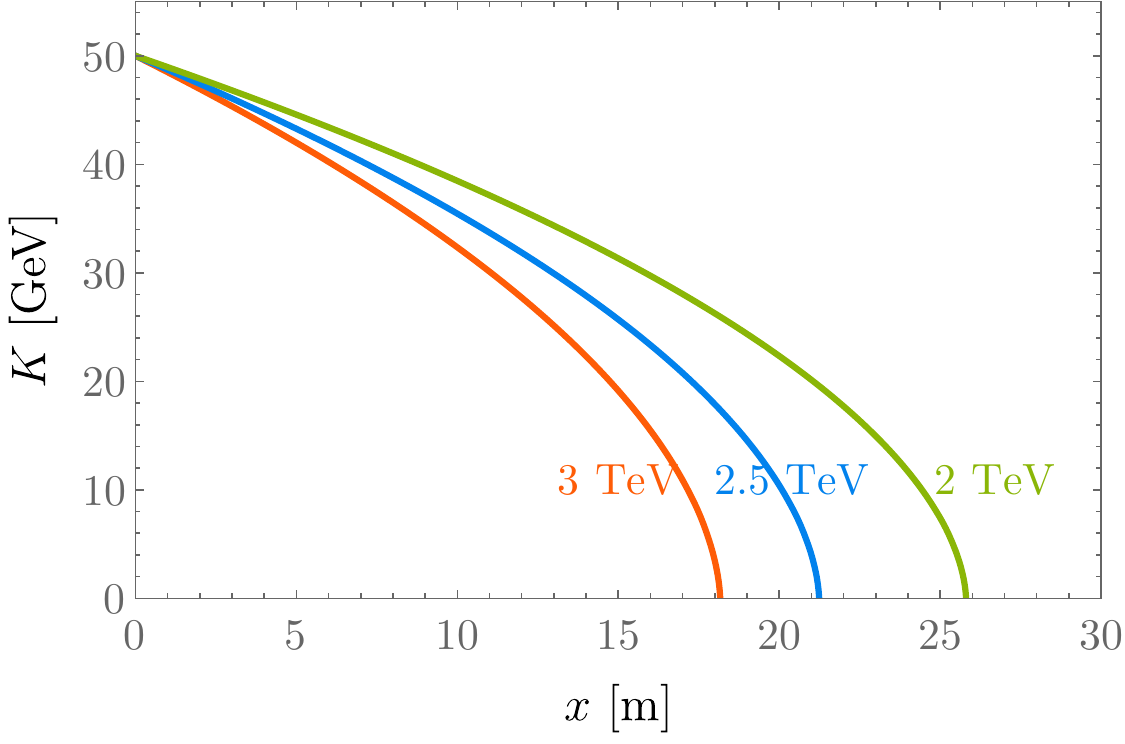}  
 \caption{Kinetic energy $K(x)$ of a gluino propagating through water as a function of distance $x$ with initial kinetic energy $K_0=50$~GeV, for various gluino masses.}
      \label{fig: gluino deceleration in water} 
\end{figure}

In Appendix~\ref{sec: dEdx numerics}, we describe our numerical solution of the $dv/dx$ equation as a charged particle traverses multiple layers of material. We show an example of such numerical solutions in Fig.~\ref{fig: gluino deceleration in silicon} by plotting the velocity $v(x)$ of a 2.5 TeV gluino with initial velocity $v_0 = 0.06$ traversing through 20~cm of silicon. The point at which the Bethe-Bloch regime transitions to the Fermi-Teller regime is labeled ``ionmax." Notice that the deceleration becomes much more efficient toward the end of the Bethe-Bloch regime, because $dv/dx \propto v^{-3}$ in the nonrelativistic limit [\eq{\ref{eq: dvdx}}].

As another example, in Fig.~\ref{fig: gluino deceleration in water}, we plot the kinetic energy $K(x)=\frac{1}{2}\mg v^2(x)$ of a gluino traversing through 30 meters of water, for various gluino masses. Since gluinos are produced with mass-independent kinetic energy distribution [Fig.~\ref{fig: E-m Distribution}], we initialize each mass at the same kinetic energy $K_0 = 50$~GeV. The dependence of the stopping distance $\xstop$ on gluino mass $\mg$ can be understood by dropping the log term in \eq{\ref{eq: dvdx}} and integrating, which gives
\begin{align}\label{eq: stopping length vs mass}
    \xstop &\approx \frac{1}{\mg} \frac{ K_0 ^2 }{2\xi C\kappa} ~.
\end{align}
Heavier gluinos therefore stop in a shorter distance, inversely proportional to their masses.

Appendix~\ref{sec: dEdx numerics} presents an algorithm that rapidly determines, for any material layer, the stopping region $\delta \bold{v}_{\text{lab}}^{\text{stopped}}$ in velocity space; we then identify the stopped gluino population by selecting simulated gluinos with $\vbold_{\text{lab}}\in \delta \bold{v}_{\text{lab}}^{\text{stopped}}$.
The simulated stopped fraction is augmented by the ratio of the Sommerfeld enhancement factors of the stopped versus total population:
\begin{align} \label{eq: Fstop}
    F_{\text{stopped}} \equiv \frac{\bar{N}_{\text{sim,stopped}}}{\bar{N}_{\text{sim,total}}}
    = \Estop \frac{N_{\text{sim,stopped}}}{N_{\text{sim,total}}} ~,
\end{align}
with
\begin{align} 
    \Estop \equiv \frac{E_S[f_{\text{stopped}}]}{E_S[f]} ~,
\end{align}
where $E_S[f]$ is defined in \eq{\ref{eq: Nsim bar}}, $f_{\text{stopped}}(\vcm)$ is the CM velocity distribution of the gluino population which satisfies $\vbold_{\text{lab}} \in \delta \bold{v}_{\text{lab}}^{\text{stopped}}$. The stopped fraction $F_{\text{stopped}}$ is ultimately used to set a limit in \eq{\ref{eq: cross section limit}}.

\begin{figure}[t] 
     \centering
    \includegraphics[width=0.492\textwidth]{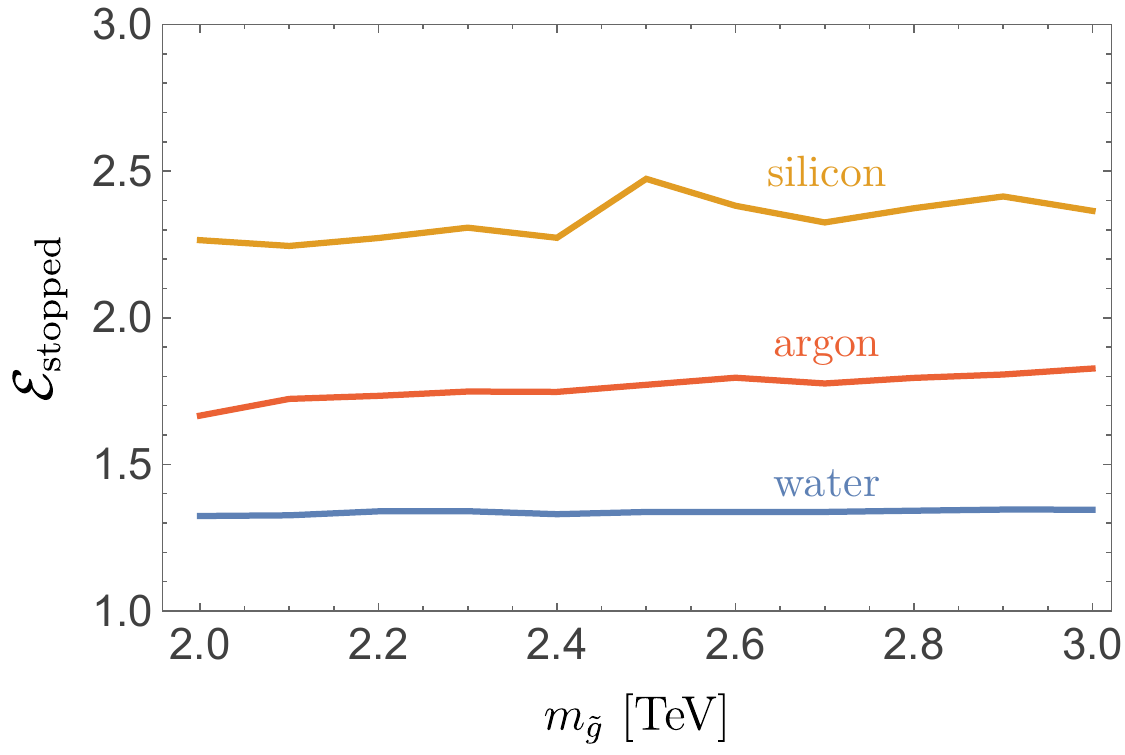}
\caption{Sommerfeld ratio $\Estop$ as a function of gluino mass $m_{\tilde{g}}$ for various velocity ranges that roughly correspond to stopping in the primary ATLAS targets.}
      \label{fig: Sommerfeld Ratio} 
\end{figure}

In Fig.~\ref{fig: Sommerfeld Ratio} we show sample Sommerfeld ratios $\Estop$  versus $\mg$ for the primary ATLAS targets, using velocity ranges that roughly correspond to stopping in silicon detectors ($v<0.06$), the EM calorimeter ($0.08<v<0.10$), and the external water pool ($0.11<v<0.24$).\footnote{As a proxy for the more complex simulation in Sec.~\ref{sec: target materials}, we apply speed-only cuts and omit angular cuts.} In all cases, $\Estop$ shows little dependence on $\mg$.

\subsection{Nuclear capture}
\label{sec: nuclear capture}
Sufficiently slow R-hadrons are likely to be captured by nuclei through their couplings with pions. The threshold velocity for stopping depends on the mass number A of the surrounding nuclei and is given by~\cite{stopping_gluino}
\begin{align}
   v_{\text{thr}} = A^{-2/3} v_F ~,
\end{align}
where $v_F \approx \Lambda_{\text{QCD}} / m_N \approx 0.25$ is the Fermi velocity of nucleons within the nucleus~\cite{nuclei_structure_book}.\footnote{Refs.~\cite{stopping_gluino,ATLAS_Rhadron_Simulation_2019} both use $v_F=0.15$ because they relate the velocity to the nucleon binding energy $\sim 8$~MeV, rather than to their kinetic energy $\sim 33$~MeV, though the latter is the more physically relevant quantity.} Due to the low rate of nuclear scattering, it is very likely that the particle stops completely due to electromagnetic deceleration before binding to a nucleus, as was assumed previously in Sec.~\ref{sec: EM deceleration}. The possibility of an abrupt, premature stopping due to nuclear binding only makes a negligible correction to the stopping length,\footnote{In contrast, the simulation by ATLAS~\cite{ATLAS_Rhadron_Simulation_2019} assumed the gluino stops abruptly when its velocity drops below the nuclear stopping threshold; this underestimates the stopping length, as shown in Fig.~\ref{fig: gluino deceleration in silicon}. In particular, the stopped particles decay searches~\cite{stopped_gluino_decay_ATLAS_2021} rely on the simulation in Ref.~\cite{ATLAS_Rhadron_Simulation_2019}, and their stopping condition could overestimate the number of stopped particles.} as we demonstrate in Appendix~\ref{sec: nuclear stopping length}.

Once captured, the R-hadron and the host atom form a TeV-scale, anomalously heavy atom.
The resultant heavy atom could behave like an isotope of the original atom if the R-hadron is neutral or could behave like a neighboring atom with $Z_{\text{eff}}=Z\pm 1$. The chemistry of this atom is primarily determined by the valence electrons of this new atom and we expect them to be near identical to a standard model atom with $Z=Z_{\rm eff}$. Owing to this similarity to known atoms, it is unlikely that the heavy atom precipitates out of the liquid or sticks to the walls of the container.
The key experimental signature is thus a small concentration of anomalously heavy atoms\footnote{In what follows, we use the terms anomalously heavy atoms, gluinos, and R-hadrons interchangeably, wherever the context permits.} embedded within the stopping material. More generally, any HSCP would result in the same class of final states; gluinos are merely considered here as a particularly well-motivated example in the context of split SUSY.

\subsection{Target materials}
\label{sec: target materials}
We consider various target materials within or near ATLAS, particularly the silicon detectors, the liquid argon in the EM calorimeter, and a hypothetical, custom-built external water volume. These targets differ in their sensitivity per unit volume, total sensitivity, and accessibility. While we focus on ATLAS, most of the discussion applies to CMS as well, except that CMS does not use liquid argon~\cite{EM_Calorimeter_CMS_vs_ATLAS_2013}.

\begin{figure}[t] 
     \centering
    \includegraphics[width=0.492\textwidth]{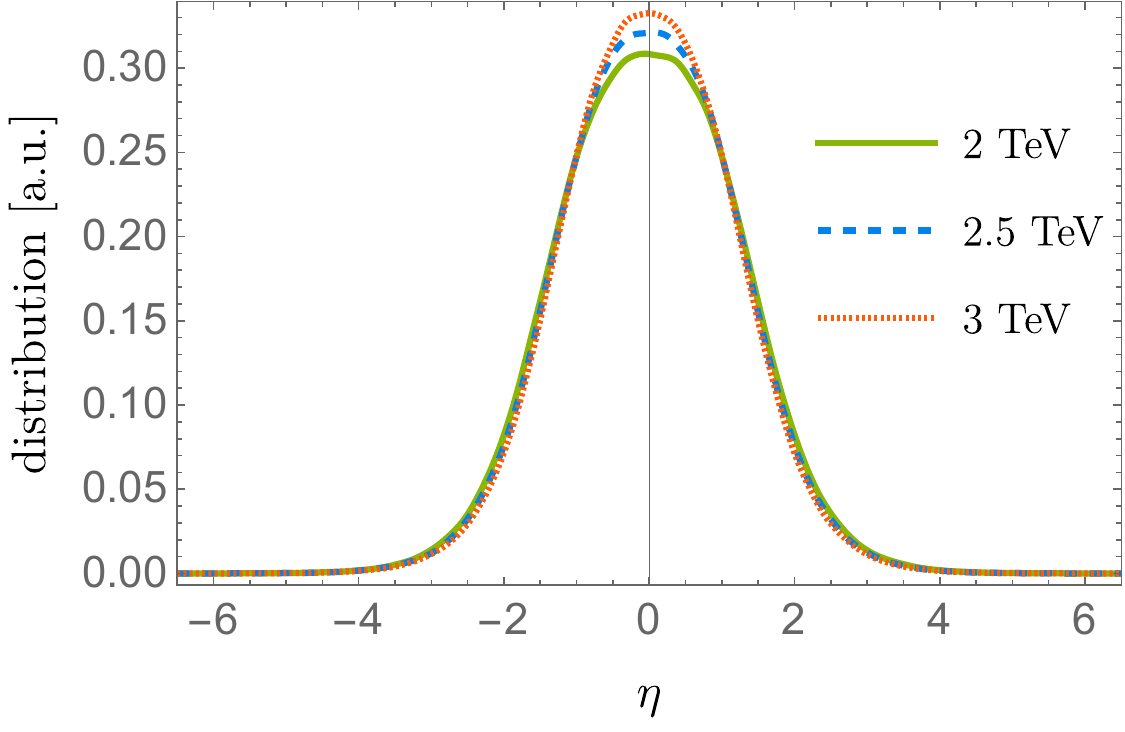}
\caption{Distribution of pseudorapidity $\eta$ for various gluino masses. The concentration of events at low pseudorapidity ($|\eta|\lesssim 1.5$) motivates the restriction of the analysis to the barrel region.}
      \label{fig: pseudorapidity} 
\end{figure}

We begin by identifying the relevant angular region. Heavy gluinos are typically produced in the transverse direction at low pseudorapidity\footnote{We adopt ATLAS coordinates with the $z$ axis along the beam line and the interaction point at the origin. The polar and azimuthal angles are $\theta$ and $\varphi$, respectively.  Directions are also expressed using pseudorapidity $\eta$, defined in terms of the polar angle $\theta$ as $\eta = - \log (\tan(\theta/2))$.} $\eta$, as opposed to the forward direction of light particles~\cite{faser_2017}. Figure~\ref{fig: pseudorapidity} shows that the $\eta$ distribution has a full width at half maximum near $|\eta| \approx 1.5$, which is well covered by the barrels of both the EM calorimeter~\cite{ATLAS_EM_Calorimeter_2013} and the silicon detectors~\cite{ATLAS_ID_Run2, ITk_Details_ATLAS_2024}. Therefore, we conservatively restrict our analysis to the barrels and neglect the small contributions from the forward endcaps.

To a good approximation, all the detector materials in the ATLAS barrels are concentric cylinders with the cylindrical axis $z$ coinciding with the beam line. The numerical simulation can thus be simplified by exploiting the azimuthal symmetry around $\varphi$, in addition to the reflection symmetry across the plane $z=0$. We apply the stopping algorithm described in Appendix~\ref{sec: dEdx numerics} but only need to make cuts in the polar angle $\theta$, effectively integrating over all $\varphi$. For each cylinder, we identify its inner radius $R$, radial thickness $\Delta R$, and axial half length $\Delta Z$ (extending from $z=-\Delta Z$ to $z=+\Delta Z$), as well as its density and other material properties. The detector materials used in the simulation are tabulated in Table~\ref{tab:material}.

For each polar-angle bin $\theta$, we trace a straight line from the origin through all cylinders whose axial half length $\Delta Z$ intersects the trajectory. The path length through each cylinder is then given by 
\begin{align} \label{eq: xf(theta)}
x_f(\theta) = \frac{\Delta R}{\sin\theta}~,    
\end{align}
where smaller values of $\theta$ correspond to trajectories more parallel to the cylindrical axis, resulting in a longer traversal along the cylinder. An angular resolution of $\delta \left(\cos \theta \right) = 0.1$ was used in the barrel simulation, which is sufficiently small for the result to converge.

\subsubsection{Silicon detectors}
We now describe the simulated detector components. The silicon detectors used in Runs 2/3, also known as the inner detector (ID), differ from the design planned for the HL-LHC, also known as the inner tracker (ITk); we will start with the ID and discuss the HL-LHC changes below.

In order of increasing radius from the interaction point, beginning with the ID system~\cite{ATLAS_ID_Run2}, two silicon subsystems are most relevant: the pixel detector and the semiconductor tracker (SCT). The barrel pixel detector comprises 4 layers, each with radial thickness $\Delta R \approx 20$~mm. The SCT also consists of 4 layers, each with thickness $\Delta R \approx 30$~mm. Between the pixel detector and the SCT is the pixel support frame (PSF), and just beyond the SCT is the transition radiation tracker (TRT)~\cite{ATLAS_ID_Run2}, both primarily made of carbon. We refer to the pixel detector and the SCT collectively as the ``silicon detectors,'' one of our primary targets. 

The silicon detectors receive the highest gluino flux because of their proximity to the collision point, resulting in high sensitivity per unit volume. The total silicon volume is approximately $V \approx 480$~L. An additional practical advantage is that the large radiation damage requires these detectors to be periodically replaced~\cite{Atlas_ITk_replacement}, conveniently allowing for our experimental analysis. A drawback is that silicon is solid at room temperature and must be liquefied and chemically processed before it can be subjected to our experimental analysis (Sec.~\ref{sec: liquefaction}). Nevertheless, analyzing retired silicon detectors is likely the most immediately accessible option.

Beyond the ID lies the central solenoid, within the barrel cryostat, whose main subcomponents are the warm and cold vessels, a support cylinder, and a thermal radiation shield~\cite{ATLAS_Solenoid}. These components are mostly made of metals such as aluminum, iron, and copper, along with carbon-based compounds. The total radial thickness of these support structures in our simulation is $\Delta R \approx 147$~mm. We included these support materials in the simulation wherever we could find their documentation, though additional components may exist that are not easily found in published reports. For simplicity, all materials are modeled as concentric cylinders; in cases such as the TRT, which comprises many small straw tubes, we approximate it as an effective cylindrical shell that preserves the original total solid volume. Additional coarse-grained approximations are employed. Further details are provided in Appendix~\ref{sec: simulated materials}.

In the HL-LHC, the ID system will be replaced by the all-silicon ITk system~\cite{ITk_Details_ATLAS_2024}, which disposes of the TRT in favor of more silicon. The ITk consists of five layers of silicon pixel detectors, with $\Delta R=$10--15~mm, and four layers of silicon strip detectors, with $\Delta R=$25~mm. It has a significantly larger total silicon volume, $V \approx 1200$~L, than the ID. Naively, this suggests greater sensitivity as there is more silicon to trap gluinos. However, the additional silicon is predominantly axial rather than radial, meaning the layers are longer rather than thicker. Because heavy particles such as TeV-scale gluinos are produced largely in the transverse direction (or low pseudorapidity; see Fig.~\ref{fig: pseudorapidity}), the radial path length through silicon at small $|\eta|$ is not significantly increased compared to Run 2. For completeness, we distinguished between the ID and the ITk in the simulations, using the appropriate configuration for Run 2/3 and the HL-LHC, respectively.

\subsubsection{Liquid argon}
Our second target is the liquid argon in the EM calorimeter, which lies just beyond the central solenoid. Its main advantage is that it is already in a liquid phase, provided the cryogenic temperature is maintained, making it straightforward to prepare for our experimental analysis. The main drawback is access: the argon is rarely removed, so a study would likely have to wait for a major detector upgrade or the end of the detector’s operational lifetime.

The active region of the EM calorimeter barrel has dimensions $R = 1500$~mm, $\Delta R = 470$~mm, and $\Delta Z = 3200$~mm, covering pseudorapidity up to $|\eta|<1.475$~\cite{ATLAS_Overview_2008}. It consists of alternating layers of liquid argon (each $\sim 4$~mm thick) and lead (each $\sim 2$~mm thick) arranged in an accordion geometry~\cite{ATLAS_EM_Calorimeter_2013}. Thus liquid argon constitutes about 2/3 of its volume, corresponding to\footnote{The experiment contains \(4.5\times10^{4}\)~L of liquid argon~\cite{ATLAS_LAr_volume}, including liquid outside the active region; the value quoted here is for the active barrel volume relevant to trapping stopped particles.} $V \approx 2\times 10^4$~L.

In our simulations we ignored the accordion structure for simplicity and instead modeled the layers as parallel to the cylindrical axis, while preserving the total volume and overall dimensions. Since the slanted geometry of the accordion structure does not significantly alter the relative path lengths through argon and lead, we do not expect it to affect the stopping power or the distribution of stopped gluinos between the two materials. Finally, rather than simulating the $\sim 80$ alternating layers of the actual geometry, we use 10 effective layers; this is sufficient for the ratio of their stopped fractions to converge to $F^{\text{Ar}}_{\text{stopped}}/F^{\text{Pb}}_{\text{stopped}} \approx 2.4$, consistent with their relative volumes.

\begin{figure}[t]
    \centering
    \includegraphics[width=0.492\textwidth]{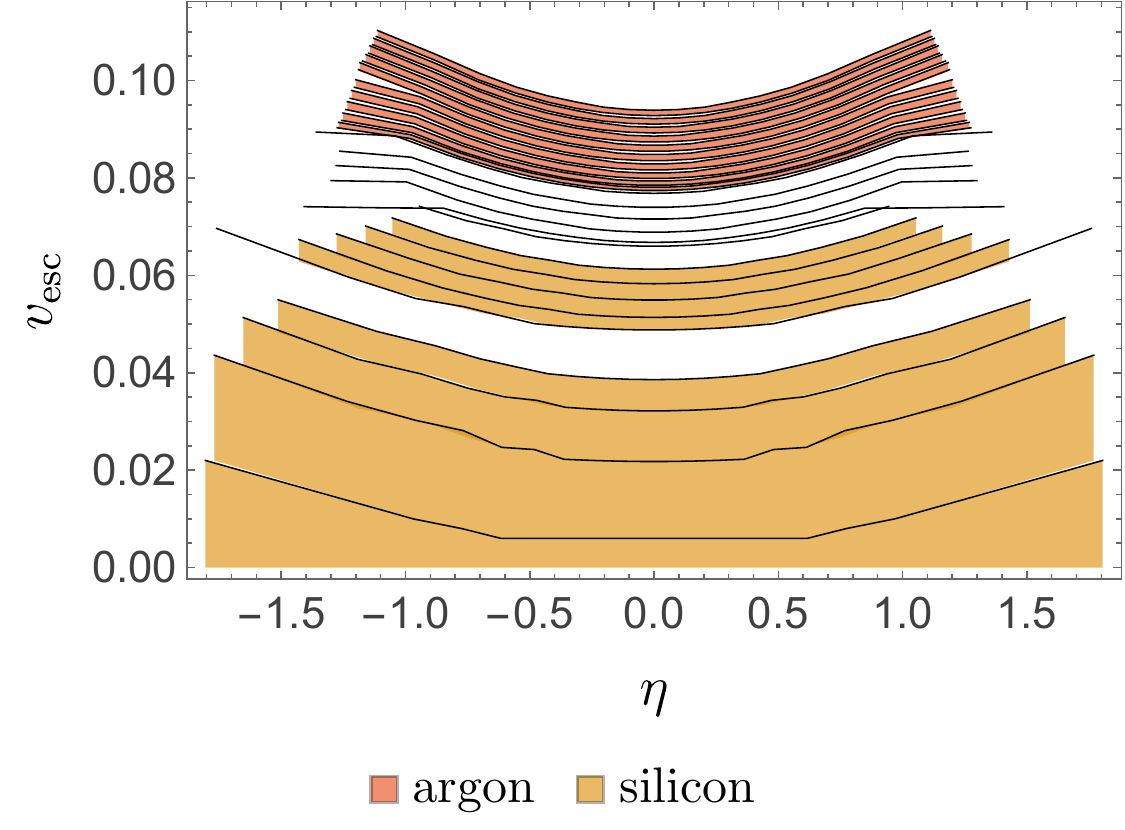}
\caption{Escape velocity \(v_{\rm esc}\) vs pseudorapidity \(\eta\) for 2.5~TeV gluinos in the ATLAS barrel for LHC Run 2/3. Here \(v_{\rm esc}\) is the maximum initial velocity for which the gluino still stops within a given layer. Black curves show \(v_{\rm esc}\) for each simulated cylindrical layer. Silicon layers in the ID and liquid argon layers in the EM calorimeter are highlighted. The liquid argon layers are separated by lead layers. See Fig.~\ref{fig: vesc vs eta HL} for corresponding result for HL-LHC.}
\label{fig: vesc vs eta Run 2}
\end{figure}

In Fig.~\ref{fig: vesc vs eta Run 2}, we show the escape velocity $\vesc$, the maximum initial velocity for which the gluino still stops within a given layer, as a function of pseudorapidity for 2.5~TeV gluinos in the ATLAS barrel. The silicon and the liquid argon layers are highlighted. The fact that the layers tend to have higher $\vesc$ at larger $|\eta|$ reflects the path length dependence on angle $x_f(\theta)$~[\eq{\ref{eq: xf(theta)}}].

Notice that the inner layers cover disproportionately larger phase space than the outer layers, reflecting a diminishing gluino capture efficiency per unit radial thickness. This behavior follows from \eq{\ref{eq: stopping length vs mass}}, which implies the escape velocity scales with the path length as $\vesc(x_f) \propto x_f^{1/4}$; since $\vesc''(x_f)<0$, each additional radial layer increases the escape speed by a smaller amount than the previous one.

\begin{figure}[t]
\centering
  \includegraphics[width=0.492\textwidth]{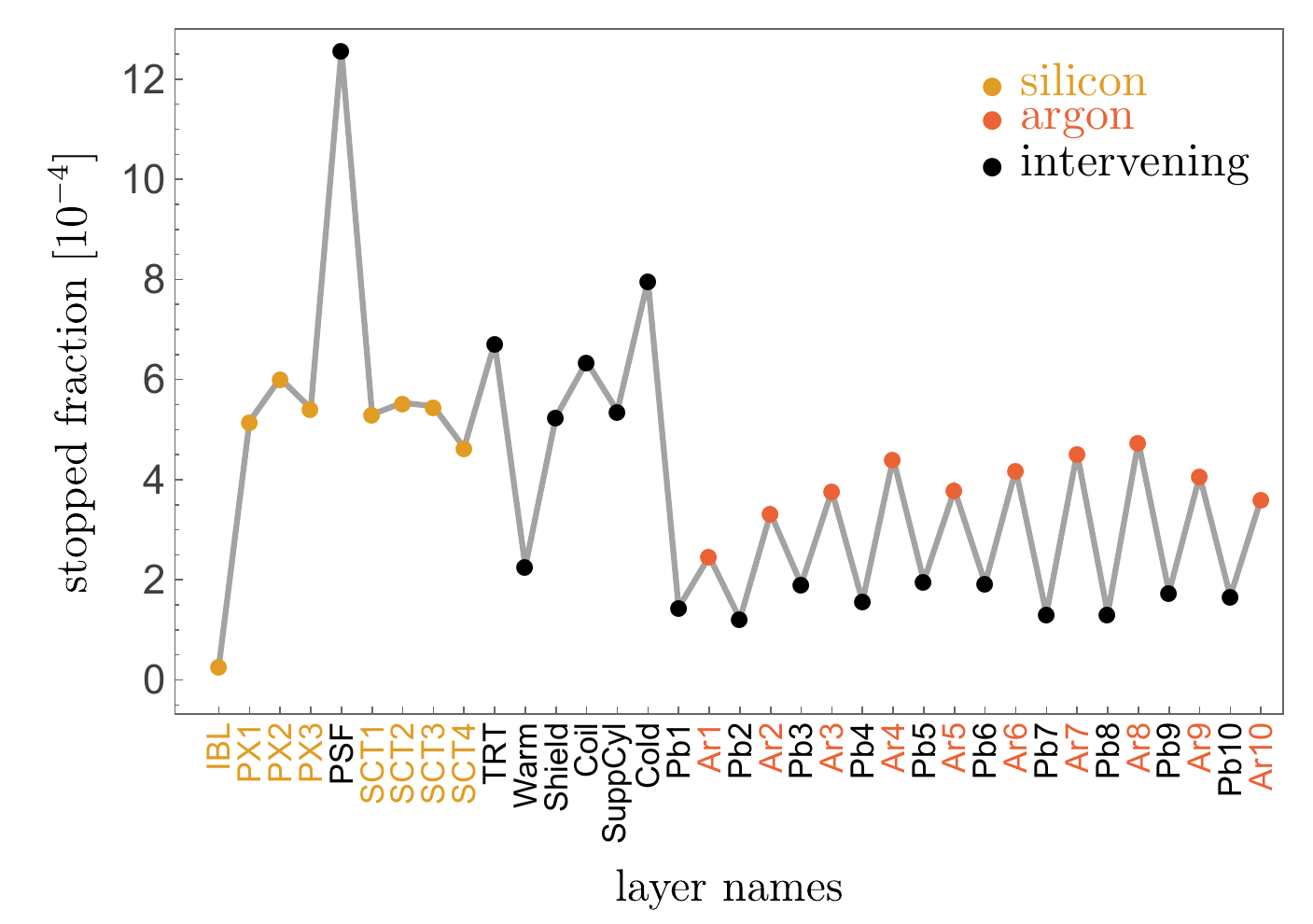}
\caption{Stopped fraction for $2.5~\text{TeV}$ gluinos in each cylindrical layer in the ATLAS barrel in Run~2/3. Silicon layers in the ID and liquid argon layers in the EM calorimeter are highlighted; all other materials in this region are labeled as ``intervening." Abbreviated layer names are shown on the horizontal axis (Table~\ref{tab:material}). The total stopped fraction of all layers shown is about 1.4\%. See Fig.~\ref{fig: stopped fraction ATLAS HLLHC} for corresponding result for HL-LHC.
}
\label{fig: stopped fraction ATLAS}
\end{figure}

Figure~\ref{fig: stopped fraction ATLAS} shows the stopped fraction for each simulated cylindrical layer in the ATLAS barrel for $\mg=2.5~\text{TeV}$, with the silicon and argon layers highlighted. These values vary only mildly across the mass range considered, as explained in Sec.~\ref{sec:projection}. 
The total stopped fraction in silicon and liquid argon are both roughly $\sim 4 \times 10^{-3}$ at $\mg = 2.5$~TeV; see Table~\ref{tab:stopped fraction}.
Since the sensitivity to gluino production cross section is directly proportional to the total stopped fraction, it is clear that analyzing both the silicon and the liquid argon improves the result by a factor of $2$. 

\subsubsection{Intervening materials}
Apart from the primary targets, a significant fraction of gluinos also stops in neighboring materials, as is evident from Fig.~\ref{fig: stopped fraction ATLAS}. In this paper, we collectively refer to all components from the collision point out to the end of the EM calorimeter, excluding silicon and liquid argon, as ``intervening material" (IM). This category includes the cryostat, the central solenoid, and the lead absorber plates of the EM calorimeter. 

While the radiation-damaged silicon detectors could be studied during operation, and the liquid argon could be extracted during long upgrades, the extraction of the intervening materials might have to wait for the end of the entire HL-LHC run; at that stage, these detector materials would have little use beyond this study. We thus consider an optimistic sensitivity scenario in which all intervening material, together with silicon and liquid argon, is liquefied and analyzed, leading to a total stopped fraction of $\sim 1.2$\% at $\mg = 2.5$~TeV; see Table~\ref{tab:stopped fraction}.

This corresponds to a threefold increase in sensitivity relative to silicon or argon alone. The total volume is $V \sim 4\times 10^4$~L, approximately twice that of the liquid argon. Since the IM consists of a wider variety of solids than silicon, chemically processing all of them would likely be more challenging, so we regard this as an optimistic target. Detector materials at even larger radii could also be considered, but a custom-built external target is likely more convenient and effective, as discussed next.

\subsubsection{Water pool}
Finally, we consider a hypothetical, custom-built volume of water outside ATLAS.\footnote{A conceptually related proposal involving water tanks near a collider to trap sleptons for subsequent decay monitoring was explored in Ref.~\cite{feng_slepton_2005}, though our method does not rely on decay signatures.} An external target offers maximal accessibility and flexibility not afforded by the internal detector targets, since the material can be analyzed at any time without interrupting collider operations. Furthermore, using water at room temperature avoids the need for liquefaction required by solids such as silicon and the need for cryogenic temperature required by liquid argon.

The main drawback of an external target is that its distance from the collision point is necessarily much greater, which reduces the gluino flux. In addition, the aforementioned diminishing gluino-capture efficiency per unit radial thickness due to the $dv/dx \propto v^{-3}$ scaling gets worse as the stopping velocity increases. Of course, these shortcomings can be mitigated by pursuing a larger water volume.

The primary constraint is the shortest available distance between the water pool and the interaction point. Along the shorter of the transverse directions to the beam line, the ATLAS cavern has a half-width of 15~m~\cite{ATLAS_cavern}; we will assume this minimal distance in the calculation below. If a substantial layer of overburden must remain intact, it may still be feasible to melt and analyze that layer of rock afterward. 

As a benchmark, we assume a cylindrical pool with radius and height $r=h=30$~m, centered at $\eta=0$, with its nearest face a distance $d=15$~m from the collision point. This geometry corresponds to pseudorapidity coverage $|\eta|<1.4$ and a volume $V \approx 10^8$~L. At this scale, the water target provides the largest total sensitivity among our three options. Of course, one could choose to construct a smaller water tank and reduce the sensitivity to a value closer to the other options, which might still have advantages in terms of ease of processing.
Although some infrastructure is required, the cost is largely limited to excavation, with no requirement on water purity and no ongoing maintenance required as it is a passive detector. Note that the geometry considered here is far from ideal in favor of practicality: the pool is assumed to be a distance $d$ next to the experiment cavern instead of surrounding it; the chosen shape of a cylinder is also just a convenient benchmark.

A detailed account of all the materials gluinos encounter in the experiment cavern prior to reaching the water pool is quite complicated. Instead, we make a coarse-grained approximation in which this entire volume is modeled as a single material (dubbed ``ATLAS material") with a uniform density $\rhoatlas$ given by the average value; the ATLAS detector is a 7000-ton cylinder with a length of 44~m and a diameter of 25~m~\cite{ATLAS_cavern}, resulting in an average density of $\rhoatlas=0.32~\rm{g}/\rm{cm}^3$. This relatively low value arises from large regions with negligible density within the detector volume~\cite{ATLAS_density_heatmap}. Since the elemental composition of the dense regions of the detector is dominated by iron~\cite{ATLAS_density_webpage}, the ATLAS material is assumed to have the same mean excitation energy $I$ as iron. We detail the simulation of the gluino traversal through the ATLAS cavern and water pool in Appendix~\ref{sec: water pool simulation}. 

\begin{figure}[t] 
     \centering
    \includegraphics[width=0.492\textwidth]{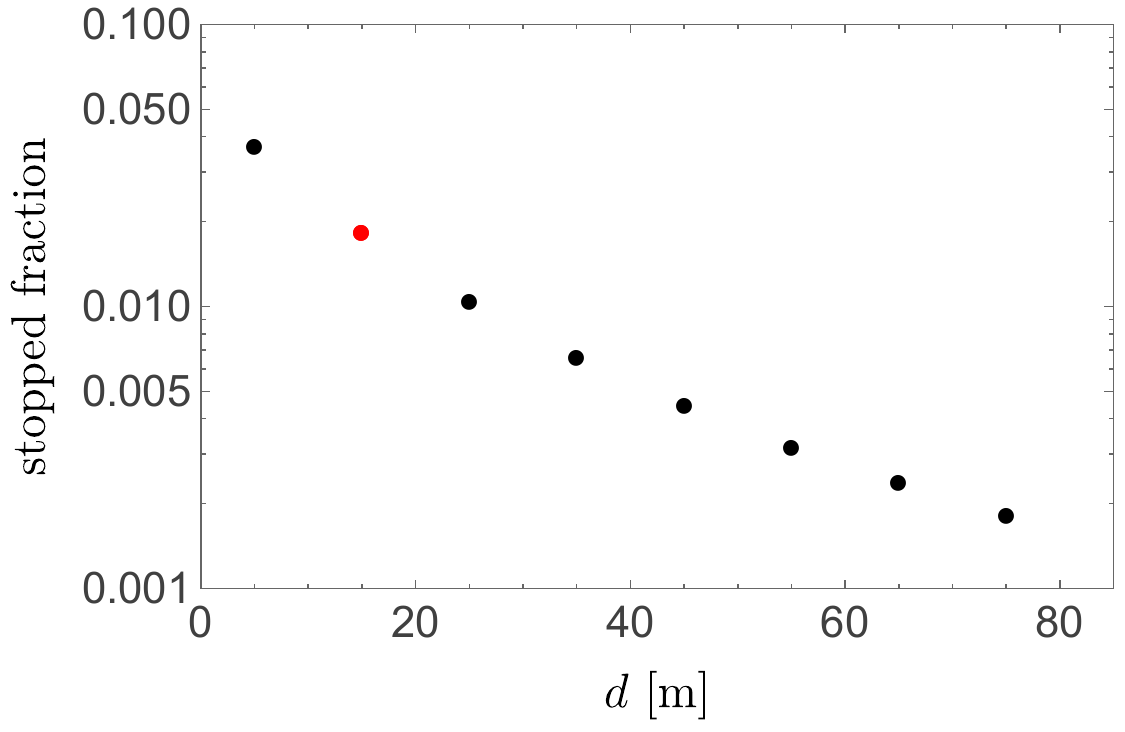}
\caption{Stopped fraction $F^{\text{water}}_{\text{stopped}}$ as a function of the distance $d$ from the collision point for a cylindrical water pool with radius and height \(r=h=30~\text{m}\) for 3~TeV gluinos. The red point is the actual distance used, $d=15$~m.}
      \label{fig: inverse square law} 
\end{figure}

Figure~\ref{fig: inverse square law} shows that the fraction of gluinos that stop in the water pool $F^{\text{water}}_{\text{stopped}}$ decreases as a function of the distance $d$ from the collision point. This is not obvious \textit{a priori} because the relevant gluino velocity range $v\sim 0.1$--0.2 lies on the rising part of the gluino lab–frame velocity distribution [Fig.~\ref{fig: vlab Distribution}]. Our result indicates that the gain from larger solid–angle coverage at smaller $d$ outweighs the modest benefit of sampling higher–velocity gluinos at larger $d$. As a rule of thumb, “closer is better” for gluino trapping, though this need not hold for particles with different velocity distributions.

For the actual ATLAS distance $d=15$~m (red point in Fig.~\ref{fig: inverse square law}), the stopped fraction is $\sim 1.8\%$ for $\mg = 3$~TeV; see Table~\ref{tab:stopped fraction}.
Since ATLAS is about $100$~m underground~\cite{LHC_Machine}, a distance of $d\sim70$--$80$~m would place the pool at the ground surface. This could substantially reduce construction cost and even enable use of possible existing surface water pools located above the experiments. Unfortunately, for gluinos the stopped fraction at such distances is an order of magnitude smaller. We leave to future work whether other hypothetical particles with velocity distributions concentrated at higher velocities could benefit from such a configuration.

With the water-pool volume and distance fixed, we checked that, within practical ranges, the dependence on the aspect ratio $a \equiv r/h$ is weak and the choice $a=1$ is near optimal. Under the volume-preserving rescaling $r \to s^{1/3} r$, $h \to s^{-2/3} h$, and $a \to s a$, we find for $\mg=3$~TeV that doubling the aspect ratio to $a = 2$ mildly increases the stopped fraction $F^{\text{water}}_{\text{stopped}} \to 1.08 F^{\text{water}}_{\text{stopped}}$, whereas reducing it to $a=1/2$ decreases $F^{\text{water}}_{\text{stopped}} \to 0.72 F^{\text{water}}_{\text{stopped}}$.

\begin{table}[t]
    \centering
    \renewcommand{\arraystretch}{1.2} 
    \begin{tabular}{c| @{\hspace{1.5em}} c @{\hspace{1.5em}} c @{\hspace{1.5em}} c}
        \hline
        \diagbox{Target}{$\mg$} & 2 TeV & 2.5 TeV & 3 TeV \\  
        \hline
        water & 1.4\% & 1.6\% & 1.8\% \\ 
        silicon & 0.26\% & 0.37\%  & 0.36\% \\ 
        argon & 0.30\% & 0.37\%  & 0.45\% \\ 
        Si+Ar+IM & 0.92\% & 1.2\% & 1.3\% \\ 
        \hline
    \end{tabular}
    \caption{Gluino stopped fraction in each target at the HL-LHC for various gluino masses. ``Si+Ar+IM" denotes combining the silicon detectors, liquid argon, and the intervening materials between them.}
    \label{tab:stopped fraction}
\end{table}

\subsection{Thermal suspension}
An important question is how stopped gluinos are distributed in matter. If the gluinos come to rest in a solid target, such as the silicon detectors, the resulting heavy atoms will be embedded within the atomic lattice. The weight of their TeV-scale mass is insufficient to break chemical bonds (see Appendix~\ref{sec:chemical bond}). Therefore, gluinos are expected to remain relatively uniformly distributed in solid.

On the other hand, if gluinos are brought to rest in a liquid target like water or liquid argon, thermal agitation ensures that the gluinos follow a Maxwell–Boltzmann distribution in height $h$~\cite{Perrin},
\begin{align}
    \rho(h) \propto e^{-\mg g h/T} \equiv e^{-h/h_0} ~,
\end{align}
where, at room temperature $T=300$ K, the scale height $h_0$ is given by
\begin{align}
    h_0 \equiv \frac{T}{\mg g} = 77~ \text{m} \left( \frac{3~\text{TeV}}{\mg}\right) \left( \frac{T}{300~\text{K}}\right) ~.
\end{align}
In other words, particles at the TeV scale are not heavy enough to settle at the bottom of normal-sized liquid containers, even at the colder temperature of $T=87$~K for the liquid argon.

Because the gluinos remain roughly uniformly distributed, achieving the target sensitivity requires processing a substantial volume of liquid, up to $\sim 10^8$ L of water. Remarkably, the efficient processing of such enormous volumes is made possible by modern centrifuge technology, as detailed in the following section.

\section{Heavy particle search in liquid}
\label{sec:heavysearch}
We describe a background-free technique to search for anomalously heavy atoms within materials from the LHC, borrowing techniques developed for heavy isotope searches~\cite{sea_water_centrifuge_proposal_1987,HeavyWater2}. The procedure consists of the following steps:
\begin{enumerate}
[label=\Alph*.]
    \item liquefaction (for solid samples),
    \item iterative centrifugation,
    \item mass spectrometry.
\end{enumerate}
Since this method requires the sample to be in liquid form, for solid samples such as silicon, liquefaction is performed as a preliminary step. The sample is then subjected to iterative centrifugation, progressively reducing its volume while retaining the heaviest constituents. Once the sample has been concentrated to a manageable volume, mass spectrometry is employed to identify trace concentrations of heavy particles. With state-of-the-art instrumentation, this procedure is, in principle, sensitive to a single anomalously heavy atom in the original sample. In what follows, we describe each step in detail and estimate achievable processing rates.

\subsection{Liquefaction}
\label{sec: liquefaction}
The liquid-phase requirement poses no difficulty for water and liquid argon, but silicon must first be chemically transformed into a compound that is liquid at room temperature. A variety of silicon-based compounds meet this criterion and can be synthesized via relatively straightforward methods. Examples include members of the silane (e.g., $\text{Si}_3\text{H}_8$), chlorosilane (e.g., $\text{HCl}_3\text{Si}$), and silicone polymer families~\cite{chemistryBook}. Among these, polydimethylsiloxane (PDMS) is widely employed in silicone oils due to its inertness, nontoxicity, and nonflammability~\cite{PDMS}. Care must be taken during the chemical processing to ensure maximal retention of the starting materials, and compound selection should account for this fabrication constraint.  Since any losses during the chemical preparation step would reduce the final reach, we emphasize that our final silicon reach is conditional on this retention efficiency, which is assumed to be lossless here. Details of the chemistry is beyond the scope of this work and left for future investigation.

Additionally, maintaining argon in liquid form requires temperatures below its boiling point of 87 K, which is colder than typical centrifuges designed for biological applications, but still above liquid nitrogen temperature of 77 K. Centrifuges maintained at liquid nitrogen temperature are used in commercial settings~\cite{drawell_refrigerated_centrifuges_2024}; even centrifuges at liquid helium temperature ($\sim 1$ K) have been demonstrated~\cite{liquid_helium_centrifuge}. Hence we do not regard this as an insurmountable difficulty.

The discussion below focuses on water at room temperature, but similar results should apply to liquid argon and silicone oil. Indeed, the much larger target volume for the water pool implies that most of the optimization constraints considered for processing time can be relaxed for targets of smaller volume.

\subsection{Iterative centrifugation}
\label{sec: iterative centrifugation}
The next step in the process is the enrichment of the liquid. In the specific example of water, we describe the procedure to reduce $\sim 10^8$ L of water down to microliter scale. Such large concentrations have been achieved in the 1970s using electrolysis for heavy water~\cite{HeavyWater1,HeavyWater2}; that method relies on differences in chemical properties which might work for our case. However, we propose a more conservative, chemistry-agnostic alternative that depends only on the mass difference. 

Therefore, we propose to enrich the liquid sample through iterative centrifugation~\cite{sea_water_centrifuge_proposal_1987}. A centrifuge separates fluid components based on their density by applying a large centrifugal force. Since our goal is to isolate TeV-scale particles from a background of normal atoms, typically $\mathcal{O}(10)$ GeV, we focus on liquid-phase laboratory centrifuges optimized for separating particles with large mass differences, as opposed to gas centrifuges, which are used for isotope separation and rely on the low viscosity of gases to resolve small mass differences~\cite{uraniumCentrifugeBook}. The most powerful class of these devices is known as ultracentrifuges.

To estimate the processing rate, consider a centrifuge with a maximum angular velocity $\omega = 2\pi f$ and a total processing volume \(\Vcen\) (i.e., the combined volume of all its test tubes). Let r denote the radial distance from the axis of rotation, and \(\rmax\) the maximum radial distance of the test tubes. A centrifuge’s performance is typically characterized by its maximum rotational frequency $f$ and its maximum relative centrifugal field (RCF), defined as \(\text{RCF} = \omega^2 \rmax / g\), where $g$ is the acceleration due to gravity. For concreteness, we adopt a commercially available ultracentrifuge as a benchmark (Eppendorf CR30NX~\cite{eppendorf}), whose parameters are listed in Table~\ref{tab:centrifugeParam}.

\begin{table}[t]
    \centering
    \renewcommand{\arraystretch}{1.2}
    \begin{tabular}{c @{\hspace{1.5em}} c}
        \hline
        RCF & $1.1 \times 10^5$ \\
        f (rpm) & $3\times 10^4$ \\
        $\rmax$ (cm) & 11 \\
        $r_0$ (cm) & 1.2 \\
        $l_0$ (cm) & 0.07 \\
        $\Vcen$ (L) & 6 \\
        $t_a$ (s) & 40~\cite{EppendorfStaff} \\
        $\teq$ (s) & 100 \\
        cost (USD) & $5 \times 10^4$~\cite{EppendorfStaff} \\
        \hline
    \end{tabular}
    \caption{Commercial centrifuge parameters for the Eppendorf CR30NX~\cite{eppendorf} as a benchmark, assuming $3$~TeV gluinos and room temperature water. RCF = relative centrifugal field, in units of gravitational acceleration $g$; $\rmax$ = maximum radial distance; $r_0$ = characteristic gluino radius; $l_0$ = gluino scale height; $\Vcen$ = total centrifuge tube volume; $t_a$ = acceleration time to reach max speed; $\teq$ = equilibration time.}
    \label{tab:centrifugeParam}
\end{table}

This centrifuge requires an acceleration time of $t_a \approx 40$~s to spin up to its maximum angular velocity~\cite{EppendorfStaff}. Following this, the system reaches sedimentation equilibrium after some additional equilibration time $\teq$, where we conservatively ignore the equilibration that started during the acceleration.

In Appendix~\ref{sec:equilibriumTime}, we estimate $\teq$ using the Drude model: to a first approximation, the gluino is diffusing through the water tube at thermal velocity, $\vth \approx 10^{-7}$, a motion that is random in direction and whose mean distance grows only as $\sim t^{1/2}$, requiring a long diffusion time $t_{\text{diff}}=5700~\text{s}$ to diffuse across the tube. Meanwhile, the centrifugal acceleration gives it a small drift velocity, $\vdrift \approx 3\times 10^{-12} \ll \vth$, in between diffusive scattering, but which is directed consistently in the outward direction; the resulting distance grows linearly in time. Modeling the scattering between the anomalously heavy atom and surrounding normal atoms as hard sphere scattering with angstrom size, and taking into account the ballistic motion of the heavy particle using its momentum-transfer cross section, we find that the equilibration time is given by the time it takes to drift across the tube, 
\begin{align}
    \teq = t_{\text{drift}}\approx 100~\text{s}\left(\frac{ 3~\text{TeV}}{\mg} \right)^{3/2}~.
\end{align}

At equilibrium, the centrifugal acceleration balances diffusion, and the radial distribution of gluinos follows the Maxwell-Boltzmann form with a potential energy $U(r) = -\mg \omega^2 r^2 / 2$ (see Appendix~\ref{sec:equilbiriumDistribution} for derivation):
\begin{align}
    \rho(r) \propto \exp\left(\frac{\mg \omega^2 r^2/2}{T} \right) \equiv \exp\left(\frac{r^2}{r_0^2} \right)  ~,
\end{align}
where $T = 300$ K is the room temperature. The positive exponent reflects the outward-pointing centrifugal force, leading to gluino accumulation at larger radii. The characteristic radius $r_0$ is given by
\begin{align}
    r_0 = \frac{1}{\omega}\sqrt{\frac{2T}{\mg}} &= 1.2~\text{cm} \left( \frac{3\times 10^4~\text{rpm}}{f}\right)\left( \frac{3~\text{TeV}}{\mg}\right)^{1/2} ~.
\end{align}

To analyze the density profile near the far end of the test tube where the gluinos accumulate, we define the coordinate $l = \rmax - r$, whose distribution is a decaying exponential, $\rho(l) \propto \exp(-l/l_0)$, where the effective ``scale height" is
\begin{align}
    l_0 =\frac{r_0^2}{2\rmax} = 0.07~\text{cm}  \left( \frac{3\times 10^4~\text{rpm}}{f}\right)^2 \left( \frac{3~\text{TeV}}{\mg}\right) ~.
\end{align}
As discussed in Appendix~\ref{sec:chemical bond}, the gluinos are not fast or heavy enough to penetrate the container walls.

To enrich the sample, we discard the fluid from the inner region of the test tube and retain only the outermost portion, defined by the retention length $l = \tilde{l} \times \rmax$, where $\tilde{l} \ll 1$ denotes a small fractional cut. This requires a setup capable of isolating the liquid at the tube’s end, a procedure we refer to as a “cut,” without inducing mixing or loss of the target liquid. One possibility is to introduce a small amount of high viscosity solution at the end of the tube~\cite{sea_water_centrifuge_proposal_1987}; we leave the exact implementation for future work. Gluino redistribution after stopping the centrifuge is not a concern due to its long diffusion timescale $t_{\text{diff}} \approx 5700$~s (Appendix~\ref{sec:equilibriumTime}).

We envision $\Ncen$ identical centrifuges operating in parallel, each processing a portion of the total material. The isolated fractions from each centrifuge are then combined into an enriched sample. Once the entire original sample has been processed and its volume reduced by a factor of $\tl$, the procedure is repeated, exponentially enriching the sample. We refer to each iteration of this procedure as a cycle.

The time required to complete the first cycle is given by
\begin{align}
t_1 = \frac{\Vsam}{\Ncen \Vcen} (t_a + \teq) ~.
\end{align}
Since the second cycle has less volume to process, it costs correspondingly less time $t_2 = \tl t_1$. More generally, the $i$th cycle takes time
\begin{align}
    t_{i} = \tilde{l}^{i-1} t_1 ~.
\end{align}
The total amount of time required for $\Ncycle$ cycles is
\begin{align}
    t_{\Ncycle} = \sum_{i=1}^{\Ncycle} t_i = t_1 \frac{1-\tilde{l}^{\Ncycle}}{1-\tilde{l}} \approx t_1~.
\end{align}
For $\tilde{l} \ll 1$, the time is completely dominated by the first cycle.
Hence, as long as we have enough resources (time and number of centrifuges) to process the sample once, we can almost arbitrarily enrich the sample using this method, with the limit being the smallest volume that we can process.
Using our benchmark centrifuge parameters, if we have $\Ncen=50$ identical centrifuges working in parallel, we can target the water pool volume of $\Vsam=0.8 \times 10^{8}$~L in about a year:
\begin{align}
\begin{split}
    &t_{\text{centrifuge}} \approx  t_1 \\ 
    &= 430
    ~\text{days} \left(\frac{50}{\Ncen}\right)  \left(\frac{6~\text{L}}{\Vcen} \right) \left(\frac{\Vsam}{0.8 \times 10^{8}~\text{L}} \right) \left( \frac{t_a+\teq}{140~\text{s}}\right)~.
\end{split}
\end{align}

Once the sample volume falls below $\Vcen = 6$ L, we can transition to using successively smaller, more precise test tubes in the final stages of the procedure. The ultimate limit is set by the smallest cut that can be made from the smallest available tube volume. We find commercially available test tubes with volumes of order $\sim 100~\mu$L~\cite{microliter_centrifuges} and conservatively take the minimum final volume to be $\Vfinal \sim 1~\mu$L. This corresponds to a millimeter-scale length, which we consider a manageable resolution for experimental control.

To minimize accumulated gluino loss over multiple cycles, the fractional retention length $\tilde{l}$ must be chosen carefully. The number of required cycles $\Ncycle$ is given by  
\begin{align}
    \tl^{\Ncycle} =  \frac{\Vfinal}{\Vsam} ~.
\end{align}
The total probability of gluino loss after $\Ncycle$ cycles is
\begin{align}
    P_{\text{loss}}(\tl) 
    &= 1- \left[ 1- \frac{\int_0^{(1-\tl)\rmax}dr \exp[(r/r_0)^2]}{\int_0^{\rmax}dr \exp[(r/r_0)^2]}\right]^{\Ncycle(\tl)} ~,
\end{align}
where $\trmax \equiv \rmax/r_0 = 9.2$ for our benchmark parameters.
This loss function is highly sensitive to the choice of $\tilde{l}$. For example, to reduce our target water volume from $\Vsam \sim 10^{8}$ L to $\Vfinal \sim 1~\mu$L, an appropriate choice would be $\tl=0.1$, for which $\Ncycle = 14$ and $P_{\text{loss}} \approx 10^{-6}$.
While this may seem overly conservative, since the retention length $l$ is more than 10 times larger than the scale height $l_0$, a more aggressive choice such as $\tl=0.01$ would increase the loss probability to $77\%$. It is therefore preferable to err on the side of more cycles, as the overall time cost is dominated by the first cycle anyway.

\subsection{Mass spectrometry}
\label{sec: mass spectrometry subsection}
After obtaining an enriched sample with a volume $\Vfinal \sim 1~\mu\text{L}$, the final step of the experiment is to identify trace concentration of TeV-scale particles with single-particle sensitivity. Many experiments of this kind using mass spectrometry (MS) or accelerator mass spectrometry (AMS) have been conducted in the context of heavy-isotope searches (reviewed in Refs.~\cite{noncollider_searches_review_2014,stable_search_review_2001,strangelets_search_review_2005}).

In MS, the sample is ionized and accelerated by an electric field to $\sim 100$~keV energy. The resulting ion beam can be mass-selected magnetically to filter out low-mass components and any remaining heavy ions can be detected as a potential signal. A measurement of the heavy ion's time-of-flight and hence mass-to-charge ratio further suppresses background~\cite{noncollider_searches_review_2014}.

A key issue here is molecular background: it is crucial to break the bonds of possible large molecules present in the liquid sample, so that any detected TeV-scale particle can be interpreted as evidence for physics beyond the standard model. Molecular fragmentation can occur during ionization (Appendix~\ref{sec: mass spec appendix}) or by passage through a thin foil~\cite{HeavyWater2}. AMS is an extension of the MS method by further accelerating the magnetically mass-selected ions to $\sim$~MeV energy using a tandem accelerator before detection, allowing for more control over molecular fragmentation~\cite{strangelets_search_review_2005}.

The sensitivity of a mass spectrometry experiment is reported in minimum detectable concentration,\footnote{Any enrichment factor attained prior to the mass spectrometry experiment (e.g. the centrifuge enrichment) is not included here.}
\begin{align}
    \cmin = \frac{N_{\text{signal,min}}}{\Nsample} \approx \frac{1}{\epsilon \Nsample} ~,
\end{align}
where \(N_{\text{signal,min}}\) is the minimum number of signal particles present in the starting sample, which contains \(\Nsample\) particles in total. The efficiency $\epsilon$ is the fraction of sample that survives the entire analysis. For the \(1~\mu\text{L}\) of enriched water considered here, \( \Nsample \approx 3 \times10^{19}\). With near-perfect efficiency $\epsilon \sim 1$, this corresponds to \(\cmin \sim \Nsample^{-1} \approx 3\times10^{-20}\). 

A classic experiment that achieved high sensitivity is Ref.~\cite{HeavyWater2} from the 1980s, which searched for a heavy hydrogen-like particle with charge $+e$ in an enriched sample of heavy water using MS.  This experiment ionized the heavy water sample into a current of $I = 8~\mu \text{A}=5 \times 10^{13}~\text{ions}/\text{s}$, which is sufficiently fast for our need: processing our $N_{\text{sample}} \approx 3\times 10^{19}$ of water molecules at this rate would take approximately a week. Reference~\cite{HeavyWater2} reached a sensitivity of $\cmin \approx 2 \times 10^{-18}$ and an efficiency of $\epsilon \approx 10^{-3}$. This is close but still falls short of the single-particle sensitivity $\epsilon \sim 1$ desired here. 
 
In Appendix~\ref{sec: mass spec appendix}, we revisit the setup of that experiment in more detail, identify the dominant limitations, and describe available improvement strategies with the goal of adapting it to our need here. Our focus on Ref.~\cite{HeavyWater2} serves only as a case study, and the final optimal design could draw on many other MS or AMS methods. In view of the various options discussed there (e.g. recycling unionized gas, efficient molecular fragmentation, magnetic mass selection that filters out all standard model atoms), achieving $\epsilon \sim 1$ appears plausible; a complete experimental design is left for future work.

As discussed in Appendix~\ref{sec: mass spec appendix}, a nondestructive ion detector could be used in the final step to allow capture of an anomalously heavy ion if a positive signal is observed; the captured particle can then be studied in detail, with the signal confirmed by arbitrarily increasing its statistical significance. This implies that our method has the unique property of having a $5\sigma$ discovery reach comparable to its projected limits. In addition, we would expect a similar signal in other detector components.

\section{Projection}
\label{sec:projection}

\begin{figure*}[t]
\centering
\begin{minipage}{0.492\textwidth}
  \centering
\includegraphics[width=\linewidth]{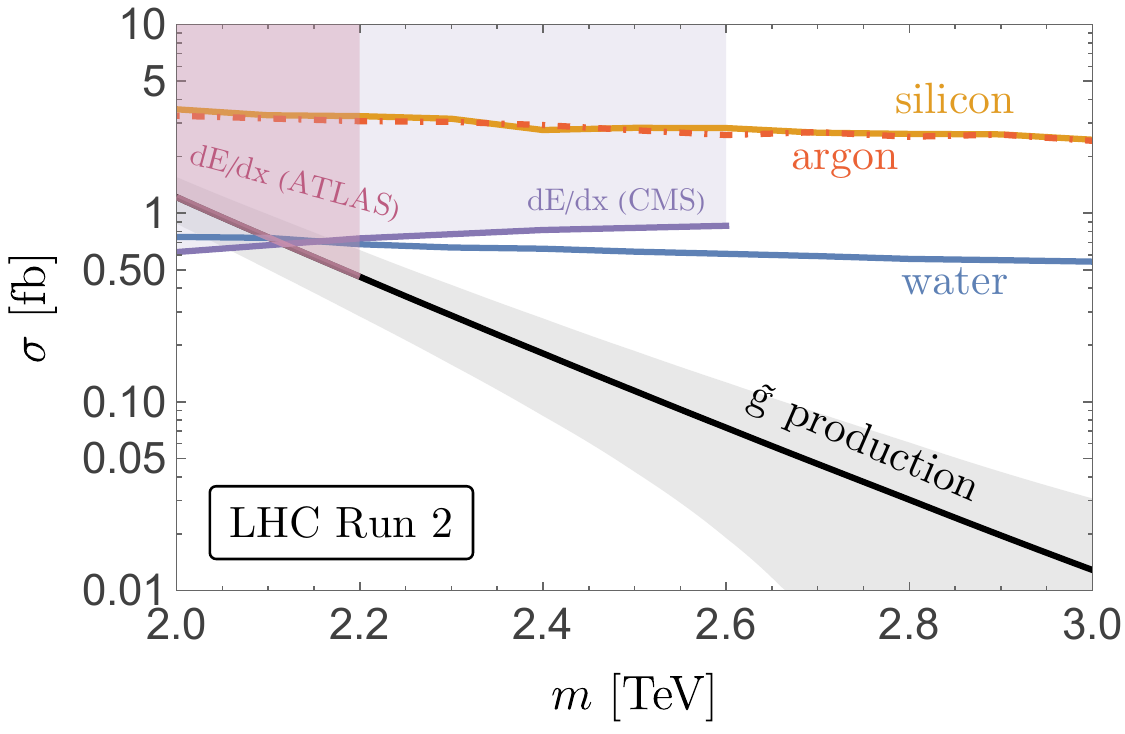}
    \subcaption{}
    \label{fig: exclusion plots Run 2}
\end{minipage}
\begin{minipage}{0.492\textwidth}
  \centering
  \includegraphics[width=\linewidth]{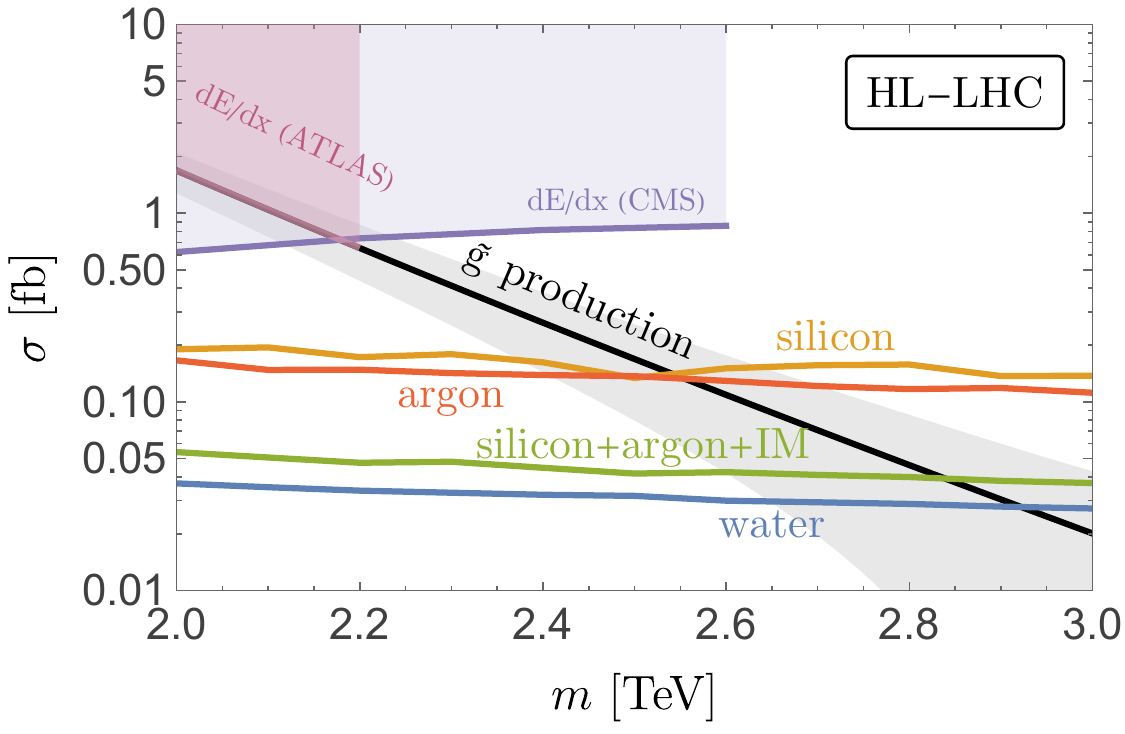}
  \subcaption{}
  \label{fig: exclusion plots HL}
\end{minipage}
\caption{Projected 95\% CL limits on gluino pair-production cross section $\sigma$ versus mass $m$ for ATLAS assuming null results in (a) LHC Run 2, $\Lint=140~\text{fb}^{-1}$ at $\sqrt{s}=13~\text{TeV}$, and (b) HL-LHC, $\Lint=3000~\text{fb}^{-1}$ at $\sqrt{s}=13.6~\text{TeV}$. Black: prediction for the gluino proper in the decoupled squarks limit, with the surrounding gray bands representing the theoretical uncertainties~\cite{gluino_cross_section_13,gluino_cross_section_13.6}. Orange: silicon in the ID barrel (Run 2) or ITk barrel (HL-LHC). Red: the liquid argon in the EM calorimeter barrel. Blue: an external cylinder of water ($r=h=30$~meters) outside the ATLAS cavern. Green: silicon, argon, and the intervening materials combined. Purple shaded regions: CMS $dE/dx$ limit ($\Lint = 101~\text{fb}^{-1}$)~\cite{CMS_dEdx_2024} and ATLAS $dE/dx$ limit~\cite{ATLAS_dEdx_2025}.}
\label{fig: exclusion plots}
\end{figure*}

The simulated gluino stopped fraction in a given target is
\begin{align} \label{eq: Fg stop}
    F_{\gt,\text{stopped}} = \frac{N_{\gt,\text{stopped}}}{N_{\gt}}~,
\end{align}
where the number of gluinos from pair production is given by
\begin{align} \label{eq: N gluino}
    N_{\gt} = 2\sigma_{\gt} \Lint ~.
\end{align}
In a more general model (denoted without the subscript $\gt$) with identical kinematics as the gluino but an arbitrary production cross section, we have
\begin{align} \label{eq: sigma ratio}
\frac{\sigma}{N_{\text{stopped}}} = \frac{\sigma_{\gt}}{N_{\gt,\text{stopped}}}  ~.
\end{align}

In a background-free experiment, a null observation in a Poisson process sets an upper limit~\cite{LeoParticlePhysicsTextBook}: 
\begin{align}N_{\text{detect}}=\epsilon_{\text{total}} N_{\text{stopped}} = -\log(1-\text{CL}) ~,
\end{align}
where $\epsilon_{\text{total}}$ is the fraction of sample that survives the entire experimental processing pipeline.
At $\text{CL}=95\%$ confidence level, this yields $N_{\text{detect}} = 3$. Substituting this into \eq{\ref{eq: sigma ratio}} and using \eq{\ref{eq: Fg stop}} and \eq{\ref{eq: N gluino}}, we obtain the 95\% CL cross section limit
\begin{align} \label{eq: cross section limit}
    \sigma = \frac{3}{2\epsilon_{\text{total}} \Lint F_{\gt,\text{stopped}}} ~.
\end{align}
In the following, we assume $\epsilon_{\text{total}} = 1$, given the protocol for achieving lossless centrifugation discussed in Sec.~\ref{sec: iterative centrifugation} and the various methods for achieving lossless mass-spectrometry discussed in Appendix~\ref{sec: mass spec appendix} and Sec.~\ref{sec: mass spectrometry subsection}.
Note that the detection of even a single trapped heavy atom translates to a discovery of an extremely high confidence level due to no backgrounds and the ability to run further tests to confirm the identity of the isolated heavy particle. Hence our discovery reach and our exclusion reach are comparable. This is in contrast to conventional observables at colliders where the 5$\sigma$ discovery reach lags the 95\% CL exclusion reach.

The resulting projection on the $\sigma$-$m$ plane is shown in Fig.~\ref{fig: exclusion plots}. Figures~\ref{fig: exclusion plots Run 2} and \ref{fig: exclusion plots HL} correspond to the Run 2 conditions ($\Lint=140~\text{fb}^{-1}$ at $\sqrt{s}=13~\text{TeV}$~\cite{ATLAS_dEdx_2025}) and the HL-LHC conditions ($\Lint=3000~\text{fb}^{-1}$ at $\sqrt{s}=13.6~\text{TeV}$~\cite{ATLAS_run3}), respectively. The theoretical gluino production cross sections are shown in black lines with surrounding gray shaded bands representing the theoretical uncertainties (which are large at the highest gluino masses); they are extracted from Refs.~\cite{gluino_cross_section_13.6,gluino_cross_section_13,table_gluino_cross_section_13.6}. The intersection between this line and an experimental projection curve sets the corresponding projected limit on gluino mass $\mg$. These limits apply for gluino lifetimes longer than the time between major collider upgrades, roughly a few years; if we continuously analyze the water or periodically extract the liquid argon, the lifetime sensitivity could be shorten to months or weeks. On the other hand, if the liquid argon or silicon need to wait until the end of the HL-LHC for processing, then the projected reach only applies to lifetime much greater than the $\sim 10$~year run time.

\begin{figure*}[t]
\centering
\begin{minipage}{0.492\textwidth}
  \centering
\includegraphics[width=\linewidth]{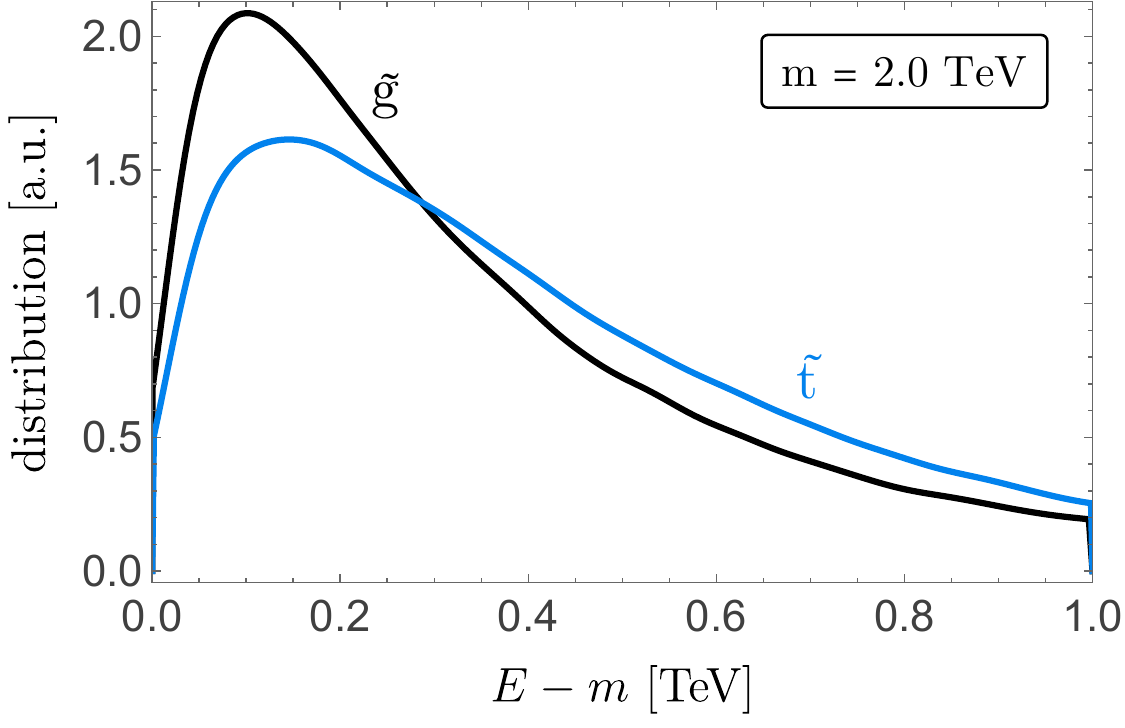}
    \subcaption{}
    \label{fig: stop vs gluino}
\end{minipage}
\begin{minipage}{0.492\textwidth}
  \centering
  \includegraphics[width=\linewidth]{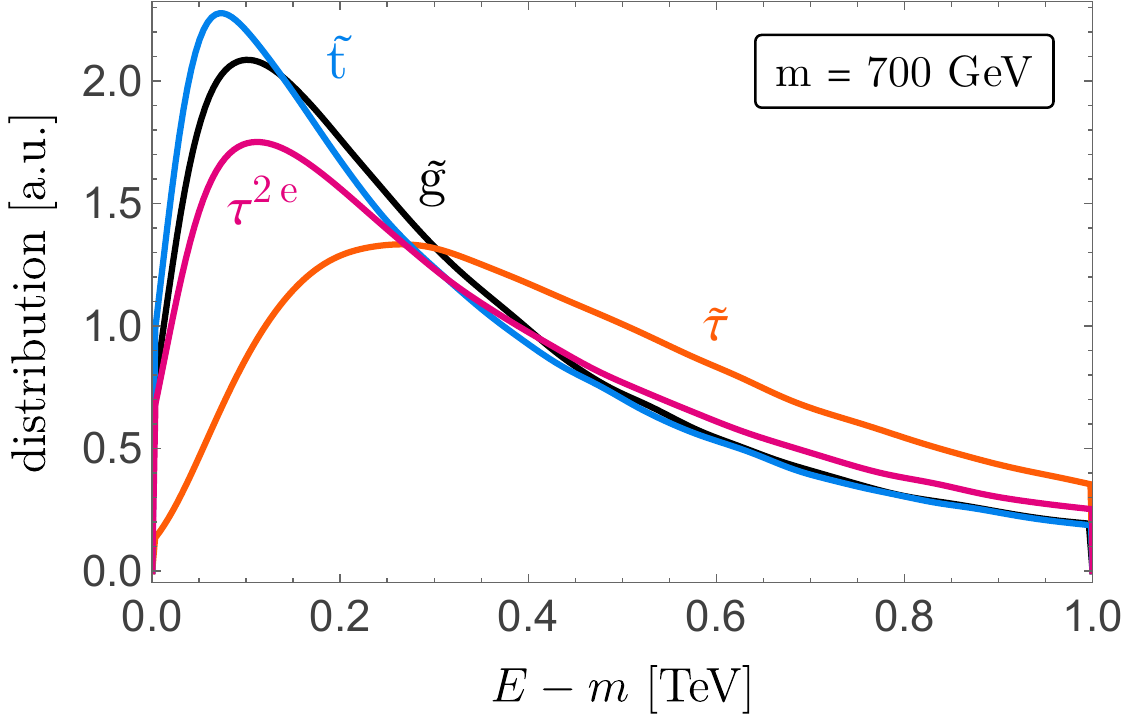}
  \subcaption{}
  \label{fig: stau vs stop vs gluino}
\end{minipage}
\caption{Comparison of kinetic energy distribution between (a) stop squarks and gluino at $m=2.0$~TeV; (b) stau and gluino and stop and doubly charged lepton at $m=700$~GeV.}
\label{fig: kinetic energy comparison}
\end{figure*}

The ATLAS $dE/dx$ search using Run 2 data with an integrated luminosity of $\Lint = 140~\text{fb}^{-1}$ sets the best current limit of $\mg = 2.20$~TeV at 95\% confidence level~\cite{ATLAS_dEdx_2025}. Since its constraint on the rest of the $\sigma$-$m$ parameter space was not reported, we make the minimal interpretation by excluding the region satisfying both $m \leq 2.20$~TeV and $\sigma \geq \sigma_{\gt}$ in Fig.~\ref{fig: exclusion plots Run 2}, shaded in red-purple. The constraint from CMS $dE/dx$ search using Run 2 data with a lower luminosity at $\Lint = 101~\text{fb}^{-1}$ is shown in the light purple curve (with the excluded region above shaded) with a limit of $\mg =2.1$~TeV~\cite{CMS_dEdx_2024}\footnote{We repeated these constrained regions in the projection plot in Fig.~\ref{fig: exclusion plots HL} for the HL-LHC for comparison. Note that these constraints only used the Run 2 data at $\sqrt{s} = 13$~TeV, and thus the intersection with the somewhat larger gluino cross section curve at $\sqrt{s} = 13.6$~TeV in Fig.~\ref{fig: exclusion plots HL} cannot be interpreted as constraints on gluino mass.}.

Though our method only stands out at high luminosity, for comparison, we also show our experimental projections assuming an identical $\Lint$ as ATLAS Run 2. Had we built the water pool near ATLAS before Run 2, our reach would have rivaled that of CMS and been slightly worse than that of ATLAS.

Interestingly, even in this low luminosity scenario, our method has an advantage compared to the $dE/dx$ method in the higher mass part of the parameter space. The $dE/dx$ method requires the use of the muon detector at the outer radius to double-check the velocity of a purported slow, heavy particle tagged earlier as a muon in the inner silicon detector. At higher mass, the velocity becomes sufficiently slow that the particle misses the muon-trigger time window, yielding a diminishing trigger efficiency at higher gluino mass and a cutoff near 2.6 TeV~\cite{CMS_dEdx_2024}, as shown by the truncated purple curve in the figure. In contrast, our method has sensitivity so long as the slow gluinos are produced.

By the end of Run 3 ($\Lint=300~\text{fb}^{-1}$ at $\sqrt{s}=13.6~\text{TeV}$~\cite{ATLAS_run3}), the projected reach of the silicon and liquid argon would each be about $\mg = 2.0$~TeV, whereas the water would reach $\mg = 2.35$~TeV (not shown in the plot).

The main advantage of our method over $dE/dx$ search is apparent at the HL-LHC in Fig.~\ref{fig: exclusion plots HL}. The silicon detectors and the liquid argon each independently have a sensitivity of $\mg = 2.55$~TeV. The external, custom-built $10^8$~L-water pool 15 meters away from the collision point reaches $\mg=2.9$~TeV. An idea appropriate for the end of the experiment lifetime is to combine silicon, argon, along with all the intervening materials (IM) between them in the barrels up to the end of the EM calorimeter, which reaches $\mg=2.8$~TeV; hence even if no water pool were built by the end of the HL-LHC, it is still not ``too late" to pursue this idea, albeit at the cost of more complicated material treatment than water. Combining ATLAS with CMS would further improve these limits.

We emphasize that our method is background-free and does not suffer from larger pile-up, hence fully realizing the benefit of an improvement in $\Lint$ at the HL-LHC. The existing $dE/dx$ search already has a lot of background~\cite{ATLAS_dEdx_2022,ATLAS_dEdx_2025}, and there is currently no projection of its performance at higher luminosity.

A notable feature of our projection curves is their weak dependence on mass. This can be understood by inverting \eq{\ref{eq: stopping length vs mass}} and solving for the initial kinetic energy $K_0$ required to stop in a material at fixed stopping length $\xstop$, which yields $K_0 \propto \sqrt{\mg}$. Because the distribution of $K_0$ is independent of $\mg$ (Fig.~\ref{fig: Kinetic Energy Distribution}), the expected ratio between the two endpoints of the mass range shown in the plot is only $\sqrt{3/2}\approx 1.22$. Small corrections to this expectation arise from the nontrivial shape of the $K_0$ distribution and from its modification by the polar angle cut, but these effects are numerically small.

\subsection{Other particles}

Although we have focused on the long-lived gluino as a benchmark, this approach likely extends the reach to other long-lived particles with a momentum distribution similar to that of the gluino. For example, we verified the stop squark ($\tilde{t}$) has a similar kinetic energy distribution [Fig.~\ref{fig: stop vs gluino}]. Since the stop also forms R-hadrons, it would undergo a similar energy loss and produce a comparable trapped population. For the water target, a production cross section sensitivity of $\sigma \sim 0.027$~fb [Fig.~\ref{fig: exclusion plots HL}] would translate to a mass of $m_{\tilde{t}} \sim 2.0$~TeV for the stop squark in the decoupled limit~\cite{gluino_cross_section_13.6}, improving upon the current bound $m_{\tilde{t}} >1.47$~TeV from the CMS $dE/dx$ search~\cite{CMS_dEdx_2024}. A detailed study of this and other particles is left for future work.

Color-neutral particles are produced with much smaller cross sections. Furthermore, they do not hadronize, but can still efficiently stop if they are electrically charged. A doubly charged lepton\footnote{We do not find much improvement over the current bounds for singly charged leptons.} ($\tau^{2e}$) could also be probed by this method with the caveat that its tremendous stopping power results in the particles stopping at much higher fraction in silicon, with smaller fractions in argon and water. Our projected limit on the $\tau^{2e}$ from silicon detector is $1.2$~TeV, improving over current bounds of 1.06~TeV by ATLAS $dE/dx$ search~\cite{multicharge_dEdx_ATLAS_2023}. The advantages of our method is likely even more pronounced for higher integer-charged leptons. 

Our method is most sensitive to particles with low kinetic energy and efficient stopping. We find that a stau ($\tilde{\tau}$) with a mass near its current limit~\cite{CMS_dEdx_2024}, $m_{\tilde{\tau}} \sim 700$~GeV, has a kinetic energy distribution peaked at $E-m_{\tilde{\tau}} \sim 250$~GeV, much higher than that of a gluino, the stop, and the $\tau^{2e}$ lepton at the same mass, which peak around $E-\mg \sim 100$~GeV [Fig.~\ref{fig: stau vs stop vs gluino}].\footnote{The differences in kinetic energy distribution can be understood as follows: The staus, being spinless scalar and produced via s-channel virtual photon or Z-boson exchange, are always produced as P-wave~\cite{slepton_production_2004, linear_collider_2001} due to angular momentum conservation; but the fermions, such as the gluinos and leptons, have spins and therefore could be produced as S-wave; the stops could also be produced as S-wave due to having t-channel production processes~\cite{stop_production_1998}.} Accordingly, our simulation shows that our reach for the stau is worse than that of the $dE/dx$ search.

\section{Conclusion}
\label{sec:conclusion}
We have proposed a novel, background-free approach to probe long-lived, TeV-scale particles, using the gluino of split supersymmetry as a representative example. In this scenario, gluinos produced at the HL-LHC could get stopped in the surroundings, and bind to standard model nuclei in detector materials or a custom-built water target. The resulting anomalously heavy atoms can later be identified through liquefaction of the material, followed by iterative centrifugation and high-sensitivity mass spectrometry.

Our simulations show that a significant fraction of gluinos produced at the HL-LHC can be stopped in existing detector materials such as silicon and liquid argon, with additional sensitivity achievable through the inclusion of intervening components or an external water target. These complementary targets can extend the discovery reach for gluino masses up to about 3~TeV, surpassing the current capabilities of conventional $dE/dx$ or time-of-flight searches while remaining insensitive to pile-up and other collider backgrounds. A similar analysis should be sensitive to stop squarks at the 2 TeV range and a doubly charged lepton up to 1.2 TeV.

While currently there are no projections for dE/dx searches at the HL-LHC for gluinos and stops, it is interesting to note that the projected reach for our method is comparable to even searches for prompt decays at the HL-LHC~\cite{SUSY_HLLHC_2014}. Furthermore, as noted earlier, the $5\sigma$ discovery reach of our method exceeds conventional searches due to the salient features of this detection method. Thus, this pushes the boundaries for the heaviest particle discoverable at the LHC.

Although our analysis focuses on the long-lived gluino as a benchmark, the same experimental strategy applies broadly to any heavy, stable or quasistable particle (e.g. the stop squark),  produced copiously in collider or beam-dump environments and subsequently trapped in surrounding materials. Our method is sensitive to lifetimes longer than a few years; with continuous sampling, this could even be shorten to months or weeks. The scalability of centrifugation and mass-spectrometric analysis to macroscopic samples, combined with the unique background-free signature, makes this a powerful and cost-effective addition to the LHC’s physics program. More broadly, it opens a new avenue for repurposing retired detector components into precision instruments for discovering new physics long after collider operations have concluded.

\acknowledgments
We thank Kaustubh Agashe, Asimina Arvanitaki, Masha Baryakhtar, David Cyncynates, Savas Dimopoulos, Isabel Garcia Garcia, Reza Ebadi, Chris Hill, Rebecca K. Leane, Kuunal Mahtani, Matt McQuinn, Martin Napetschnig, Philip Schuster, Erwin Tanin, and Natalia Toro for discussions. We thank representatives of Eppendorf and Ming Yang for helpful discussion of centrifuges.

This work was supported in part by NSF Grant No.~PHY-2310429, NSF Grant No. PHY-2515007, Simons Investigator Award No.~824870,  the Gordon and Betty Moore Foundation Grant No.~GBMF7946, the John Templeton Foundation Award No.~63595, The University of Delaware Research Foundation, and the U.S.~Department of Energy~(DOE), Office of Science, National Quantum Information Science Research Centers, Superconducting Quantum Materials and Systems Center~(SQMS) under Contract No.~DE-AC02-07CH11359.
J.~G. is supported by the U.S. Department of Energy under Contract No. DE-AC02-76SF00515. S.~R.~is supported in part by the U.S.~National Science Foundation~(NSF) under Grant No.~PHY-1818899 and  by the Simons Investigator Grant No.~827042.
S.~W. is supported in part by the U.S. Department of Energy Office of Science under Award No. DE-SC0024375, and by the Gordon and Betty Moore Foundation through Grant GBMF13898 to the University of 
Washington.

\appendix

\section{Sommerfeld Enhancement}
\label{sec: Sommerfeld}

\begin{figure}[t]
\begin{minipage}{0.492\textwidth}
  \centering
\includegraphics[width=\linewidth]{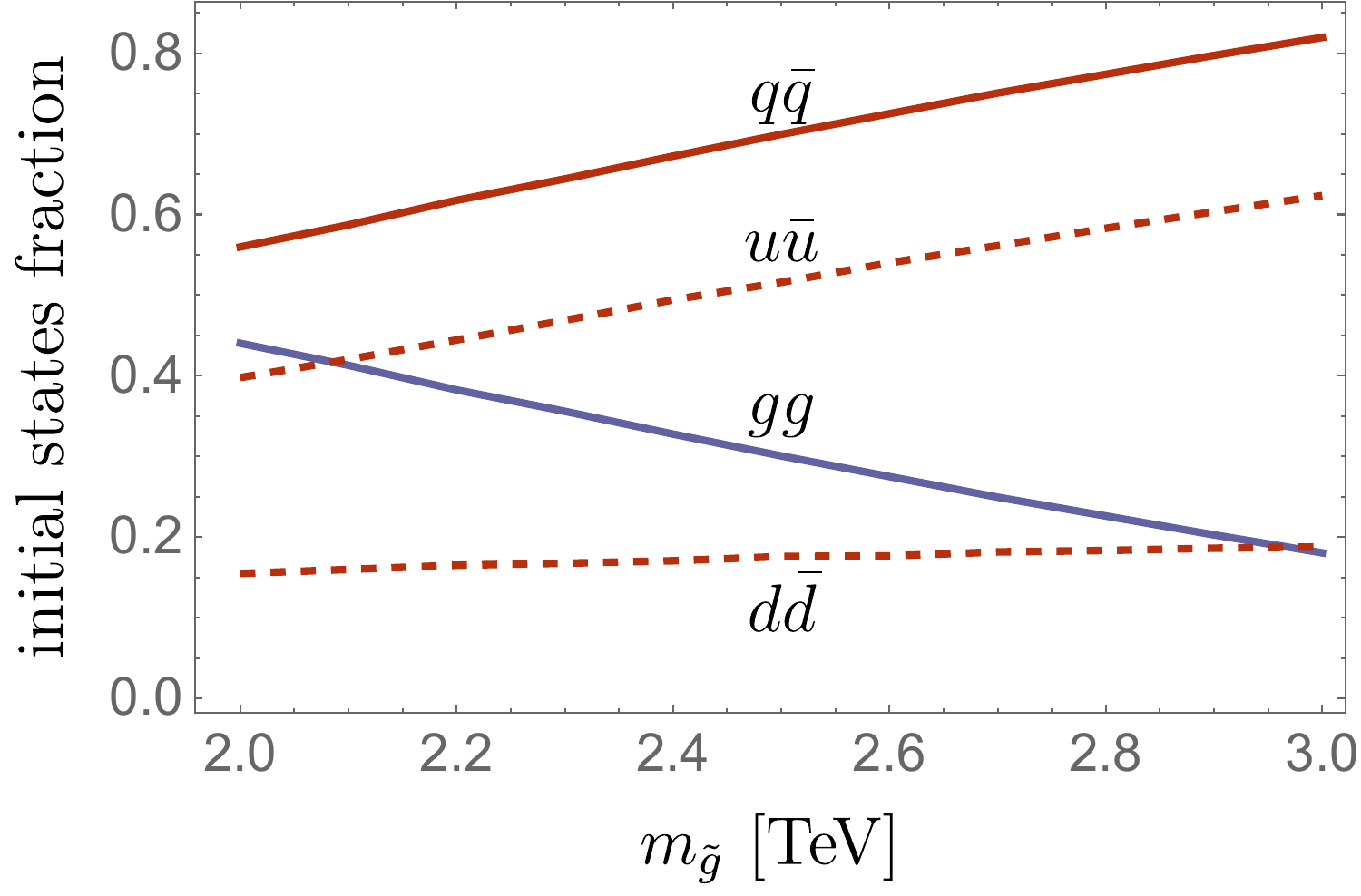}
\subcaption{}
\label{fig:parton} 
\end{minipage}

\begin{minipage}{0.492\textwidth}
  \centering
\includegraphics[width=\linewidth]{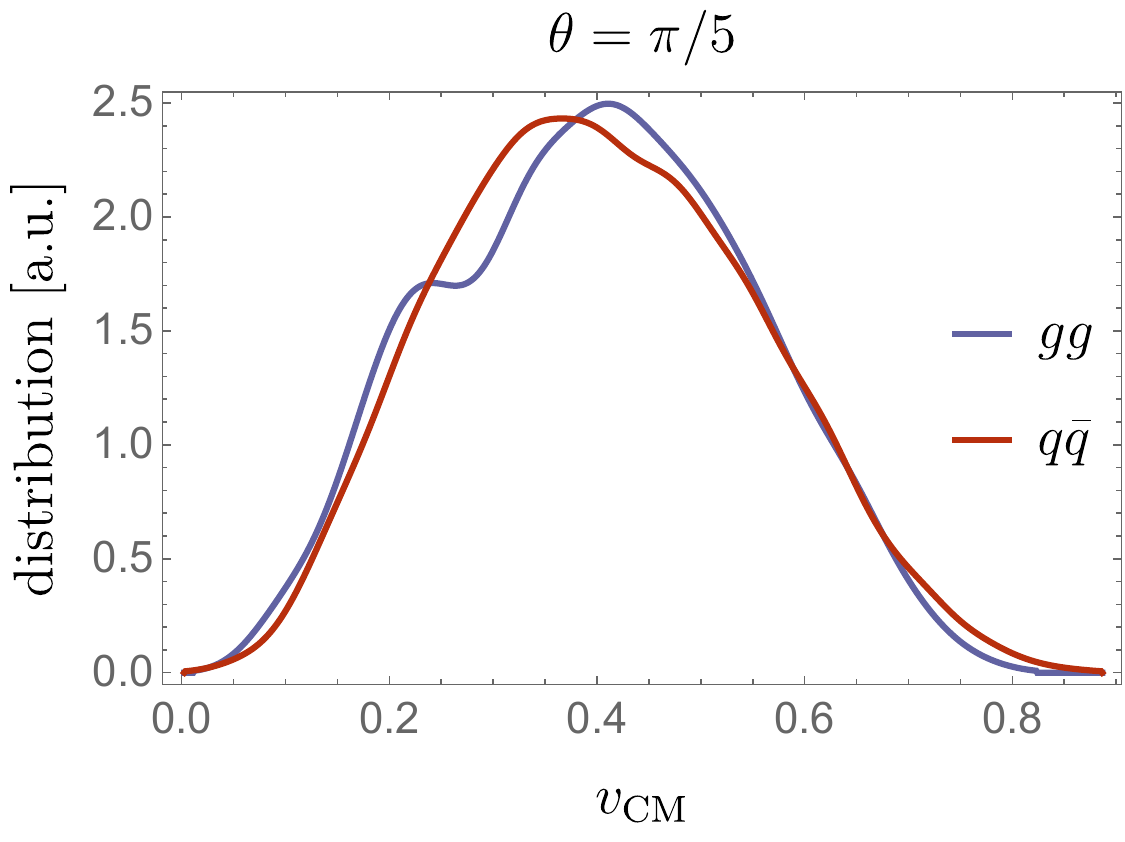}
\subcaption{}
\label{fig: velocity distribution by initial states}
\end{minipage}
\caption{(a) Contributions of various parton subprocesses for gluino production. At larger gluino masses, the initial states are dominated by quark-antiquark pairs (mostly up and down) rather than gluon pairs, giving rise to stronger Sommerfeld enhancement. Charm and strange quarks make up less than 1\% (not shown). (b) Comparison of the CM-frame velocity distributions of 2.5 TeV gluinos at the polar angle $\theta=\pi/5$ for different initial states. The solid angle bin has width $\Delta \theta = \Delta \varphi = \pi/10$.}
\label{fig: gg vs qq}
\end{figure}

The Sommerfeld enhancement factor is given by 
\begin{align}
    E_S(\vcm) = \frac{C\pi \alpha_s}{\vcm} \left[ 1- \exp\left(- \frac{C \pi \alpha_s}{\vcm} \right) \right]^{-1} ~,
\end{align}
where $\vcm = \sqrt{1-4\mg^2/\hat{s}}$~\cite{Sommerfeld_Factor}, with $\hat{s}$ being the partonic Mandelstam variable. 
Since the typical momentum transfer of the soft gluon exchanges is $Q \sim \mg \vcm$~\cite{Sommerfeld_Factor}, we take $\alpha_s(Q) \sim \alpha_s (0.1 \mg) \approx 0.1$ for \( \mg \in [2, 3] \)~TeV~\cite{PDG2024_QCD}.
The process-dependent factor $C$ is given by
\begin{align}
    C=\begin{cases}
    \dfrac{3}{2},  \quad q\bar{q}\to \tilde{g}\tilde{g}\\[1em]
    \dfrac{1}{2},   \quad gg \to \tilde{g}\tilde{g}~.
    \end{cases}
\end{align}
The $C$-value of the $gg \to \tilde{g}\tilde{g}$ process arises from a color averaging over both attractive and repulsive color final states, whereas $q\bar{q}\to \tilde{g}\tilde{g}$ always proceed through the s-channel and the final gluino pair is always in an attractive color octet state~\cite{Sommerfeld_Toolbox}. Due to the parton distribution functions, the \( q\bar{q} \) contribution increases with \( \mg \) and starts dominating over \( gg \) at the higher end of our mass range, as shown in Fig.~\ref{fig:parton}, in contrast to earlier analyses at lower \( \mg \)~\cite{stopping_gluino}.

Despite this complication from the $C$-factor’s dependence on the initial state, the velocity distribution of the final state gluinos, in both magnitude and direction, is largely independent of the initial states. In Fig.~\ref{fig: velocity distribution by initial states}, we demonstrate this for a random polar angle. This allows for a simplification in our treatment of the Sommerfeld factor, in which we take the process-dependent factor $C$ to be the weighted average between $1/2$ and $3/2$, weighted by the initial states fraction at a given gluino mass $\mg$, as in Fig.~\ref{fig:parton}.

\section{Numerical solution of $dE/dx$}
\label{sec: dEdx numerics}
Consider a particle traversing a material in a straight line. Suppose its velocity upon entry is $v_0$ and the material has length $x_{f}$ along the particle's line of motion. We determine $v(x)$ by solving the initial value problem
\begin{align} \label{eq: dvdx IVP}
    \frac{dv}{dx} = \begin{cases}
       \dfrac{(1-v^2)^{3/2}}{\mg v} \dfrac{dE}{dx}, \quad v \geq \vionmax\\ \\
         - \dfrac{2 \xi C e^{3\kappa - 1}}{3 \mg}, \quad v \leq \vionmax
    \end{cases} ; \quad v(0)=v_0 ~.
\end{align}
The particle stops within the material if it reaches $v(\xstop) = 0$ for some $\xstop \leq x_f$. If the particle does not stop, we repeat for the next material along the line of motion with initial velocity set to $v(x_{f})$.

The Fermi-Teller regime was trivially integrated analytically, while the Bethe-Bloch regime was solved numerically. We need only solve the ODE once in each setting and use the following interpolation method. For a given gluino mass and a given material type, we need to rapidly compute the final velocity as a function of initial velocity and material length, $v_f(v_0,x_f)$. We choose a pair of boundary conditions $(v_0,x_f)=(v_0^{\text{large}},x_f^{\text{large}})$ larger than all of our application's need, and solve for the corresponding function $v^{\text{large}}(x)$. Then we build a grid by sampling many examples of $v_f(v_0,x_f)$ from this function and interpolate the result.

This calculation assumes the particle moves in a straight line. The axial magnetic field with strength $B=2$~T near the collision point at ATLAS~\cite{ATLAS_Solenoid}, relevant for the silicon detectors, has negligible influence on the particle's trajectory for the mass and velocity of interest to us, as its cyclotron radius is 
\begin{align}
 R_c = \frac{\mg v}{eB} = 330~\text{m} \left(\frac{v}{0.1}\right) \left( \frac{\mg}{2~\text{TeV}} \right) ~,  
\end{align}
much larger than the relevant length scale.

At low velocities, \( v \sim 0.1 \), small deviations from the Bethe--Bloch equation arise from the Bloch correction (proportional to \( z^4 \)) and the Barkas correction (proportional to \( z^3 \))~\cite{ICRU_Stopping}. Since the gluino oscillates among electrically positive, negative, and neutral states, the Barkas correction averages to zero. We have checked that the Bloch correction modifies the stopping calculation only at the $\sim 10\%$ level, which we neglect.

Instead of simulating the propagation of every particle, which would be computationally expensive, we implement an ``escape velocity" algorithm whose computation cost is limited only by the number of target materials and their angular resolution. For every relevant material in the detector, we first determine the direction of the particles from the collision point that would pass through the material, specified by a central polar angle $(\theta_0, \varphi_0)$ and a surrounding solid angle $\delta \Omega$. Then for every polar angle bin in $\delta \Omega$, we compute the escape velocity $v_{\text{esc}}$ along the line of motion for that material; that is, a particle faster than $v_{\text{esc}}$ would not stop in this material, taking into account all previous materials encountered along the line of motion. The procedure used to compute $v_{\text{esc}}$ is described in the following paragraph. For the $k$-th material along the line of motion, $v_{\text{min}}^{\text{stopped}}=v_{\text{esc},k-1}$ and $v_{\text{max}}^{\text{stopped}}=v_{\text{esc},k}$ give the range of velocities that will stop in it. Denote the resulting phase space region capable of stopping in the material as $\delta \bold{v}_{\text{lab}}^{\text{stopped},k}$.

To determine the escape velocity, we need to find the critical initial velocity $v_0$ such that $v_f=v(x_f) \approx 0$ within some numerical tolerance, which we take to be $\vstop = 10^{-6}$. We accomplish this using a binary search algorithm: Start with the trivial statement that the true value of $v_0$ must be in the range $(v_{0,\min}=\vstop, v_{0,\max}=0.99)$, or whatever value of $v_{0,\max}$ appropriate to the problem. Then we take the median velocity $v_{\text{median}}=(v_{0,\min}+v_{0,\max})/2$, propagates it through the initial value problem of \eq{\ref{eq: dvdx IVP}} to compare it with the true value of $v_0$, allowing us to shrink the possible range in half by updating either $v_{0,\min}$ or $v_{0,\max}$ to $v_{\text{median}}$. We repeat this procedure until the possible range of $v_0$ is smaller than some threshold, which we take to be $\vstop$. This algorithm converges rapidly and takes advantage of the fact that $v_f(v_0,x_f)$ is a monotonic function of $v_0$.

\section{Nuclear stopping length}
\label{sec: nuclear stopping length}
We show that premature gluino stopping due to nuclear binding only makes a negligible correction to the stopping length. Take silicon, for instance, whose threshold velocity is $v_\text{thr} \approx 0.027$, and whose stopping distance beneath $v_\text{thr}$ due to ionization is about $\xthr \approx 3.6$~cm for a 2.5 TeV gluino; see Fig.~\ref{fig: gluino deceleration in silicon}. Since the nuclear interaction is Poisson-distributed with an average length of about $\lambdaint=47$~cm, the probability of nuclear stopping before $\xthr$ is
\begin{align}
    1 - e^{-\frac{\xthr}{\lambdaint}} \approx \xthr/\lambdaint ~.
\end{align}
On average, this occurs half way at $\xthr/2$. So the corrected stopping distance should be
\begin{align}
    x_\text{corrected} &= \xthr - \left(\frac{\xthr}{\lambdaint}\right)\left(\frac{\xthr}{2} \right)\\
    &= \xthr \left(1-\frac{\xthr}{2\lambdaint} \right) ~,
\end{align}
which works out to $x_\text{corrected} = 0.96 \xthr$ for the example here.
We ignore this small reduction in $\xthr$, which is itself a small fraction of the full stopping length $\xstop$ from the initial velocity.

\section{Water pool simulation}
\label{sec: water pool simulation}

\begin{figure}[t] 
     \centering
    \includegraphics[width=0.492\textwidth]{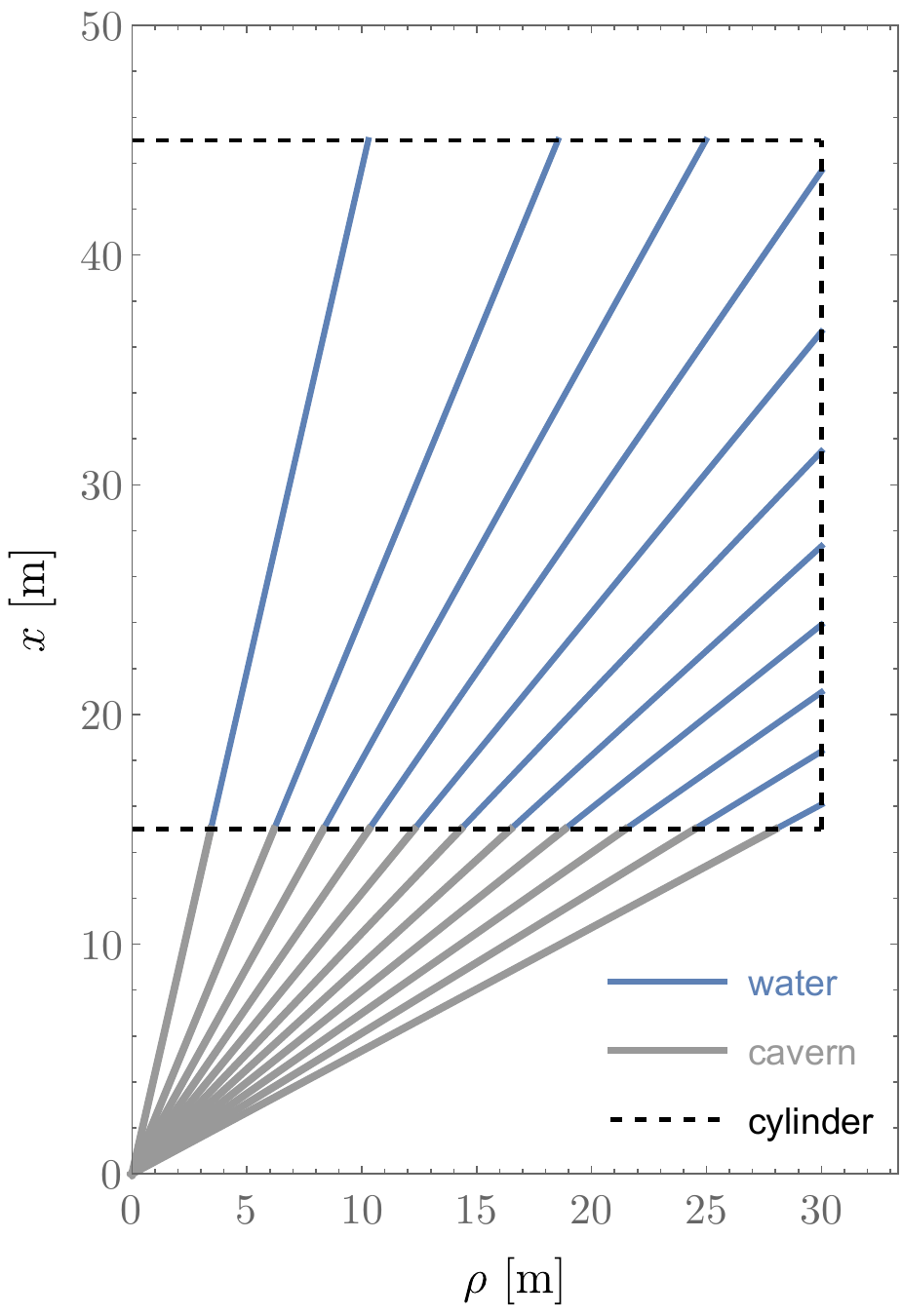}
\caption{Sampled particle trajectories in the \(\rho\)\(x\) plane through the experiment cavern (gray) and an external water cylinder (blue), with the cylinder boundary shown by the black dashed line.  Here, \(\rho\) is the radial distance in the \(yz\) plane, with the \(z\) axis along the beamline and the collision point at the origin. An implicit rotation by the azimuthal angle \(\varphi_x\) about the \(x\) axis from \(0\) to \(2\pi\) is understood.}
      \label{fig: rays through water} 
\end{figure}

To model the gluino traversal of the experiment cavern and the water cylinder, we take the interaction point to be at the origin, the beam line as the $z$ axis, and the $x$ axis to point vertically upward toward the ground surface. Define the polar angle relative to the $x$ axis as $\alpha$ (not to be confused with the polar angle $\theta$ relative to the $z$ axis); the cylinder is symmetric around the $x$ axis along its azimuthal angle $\varphi_x = 0$ to $2\pi$. The center of the cylinder's bottom face is at $(x,y,z) = (d,0,0)$. Although we have set the water pool to be above ATLAS in this picture, the system is cylindrically symmetric around the $z$ axis and there is considerable freedom in the precise location of the pool.

Consider rays emanating from the origin, representing straight-line particle trajectories (Fig.~\ref{fig: rays through water}). The rays that intersect the cylinder with radius $r$ form a cone given by $0 < \alpha < \alpha_0$, where 
\begin{align}
    \alpha_0 \equiv \arctan \left( \frac{r}{d} \right) ~.
\end{align}
Among them, an inner cone of rays that fully traverse the cylinder height $h$ is given by $0 < \alpha < \alpha_1$, where 
\begin{align}
    \alpha_1 \equiv \arctan\left( \frac{r}{d+h}\right)~.
\end{align}
Rays with intervening polar angles $\alpha_1<\alpha<\alpha_0$ hit the wall of the cylinder without reaching its farther face.

The length $\sentry$ of a ray before entering the cylinder is given by
\begin{align}
\sentry(\alpha) = \frac{d}{\cos \alpha}~; \quad 0<\alpha<\alpha_0 ~.
\end{align}
The length $\sexit$ of the ray upon exiting the cylinder depends on whether it is inside the inner cone:
\begin{align}
    \sexit(\alpha) = 
    \begin{cases}
        \dfrac{d+h}{\cos \alpha}~; \quad 0 < \alpha < \alpha_1 \\
        \\
        \dfrac{r}{\sin \alpha}~; \quad \alpha_1 < \alpha < \alpha_0 ~.
    \end{cases}
\end{align}
The cylinder traversal distance can be shown to be
\begin{align}
    \scylinder(\alpha) &= \sexit(\alpha) - \sentry (\alpha) \\
    &= \frac{1}{\cos \alpha} \min \left(h ~, \frac{r}{\tan \alpha}-d \right)
\end{align}
for all angles satisfying $0<\alpha<\alpha_0$. We thus simulated gluino trajectories (Fig.~\ref{fig: rays through water}) through ATLAS material of length $\sentry(\alpha)$ and a subsequent $\scylinder(\alpha)$ through water, using an angular resolution\footnote{At larger distances, we always sampled at least 10 angles.} of $\delta (\cos \alpha) = 0.05$.

\begin{figure}[t]
\centering
\begin{minipage}{0.492\textwidth}
  \centering
  \includegraphics[width=\linewidth]{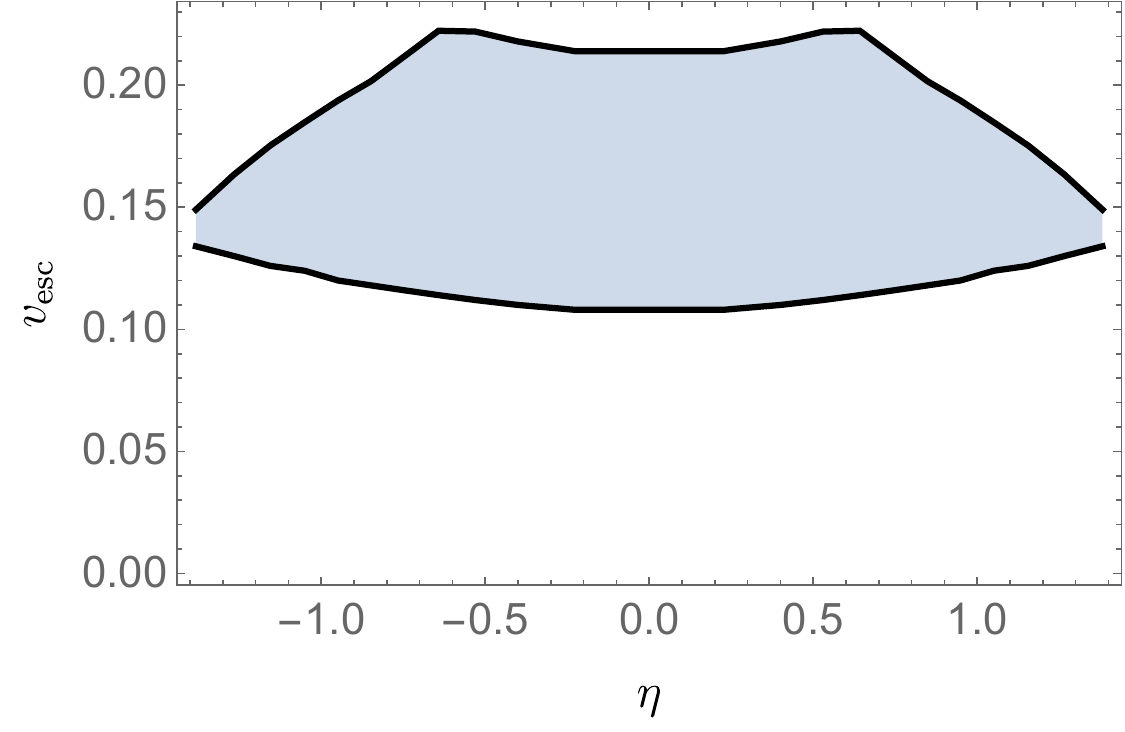}
  \subcaption{}
  \label{fig: water vesc vs eta}
\end{minipage}
\begin{minipage}{0.492\textwidth}
  \centering
  \includegraphics[width=\textwidth]{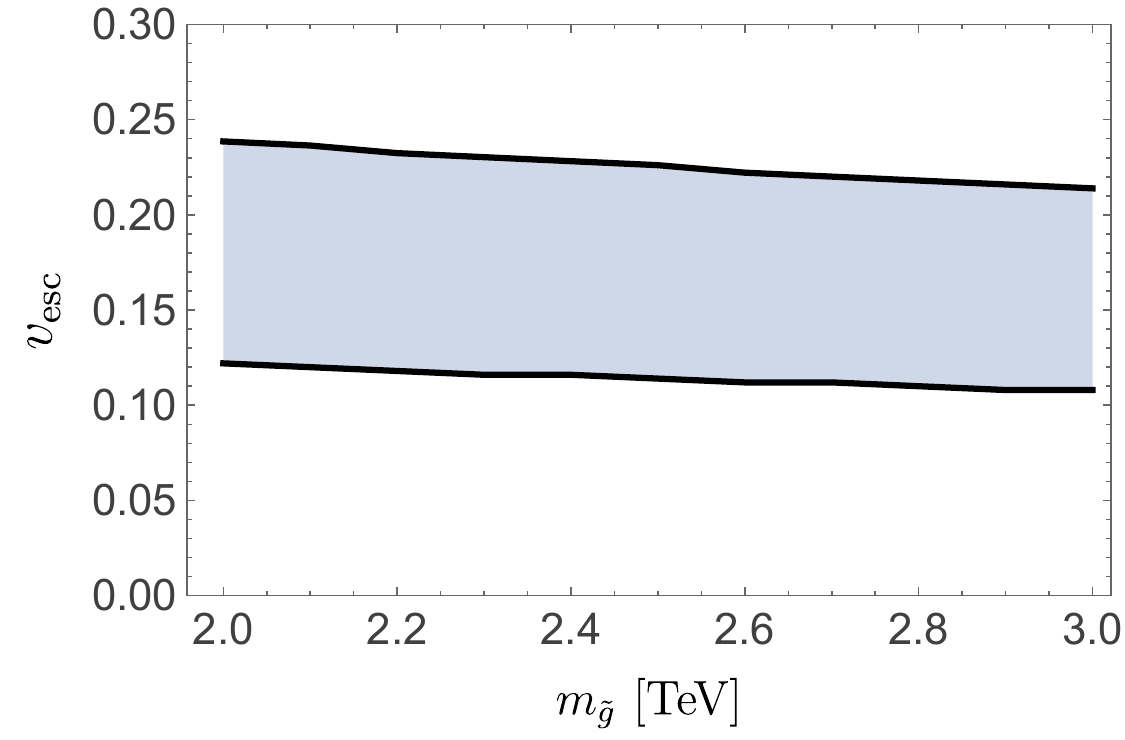}
  \subcaption{}
  \label{fig: water vesc vs mass}
\end{minipage}
\caption{Escape-velocity window for gluinos stopping in the water pool as a function of (a) pseudorapidity $\eta$ at fixed mass $\mg=3$~TeV; (b) mass \(m_{\tilde g}\) at fixed pseudorapidity $\eta=0$. The shaded band is the stopping region bounded by \(v_{\text{esc}}^{\min}\) and \(v_{\text{esc}}^{\max}\).}
\label{fig: water vesc}
\end{figure}

Figure~\ref{fig: water vesc vs eta} shows the escape-velocity window for gluinos as a function of pseudorapidity $\eta$ for \( \mg = 3\)~TeV. The kinks in the upper curve \(v_{\text{esc}}^{\max}\) at $\eta=\pm 0.63$ correspond to the critical angle $\alpha_1$, beyond which the cylinder traversal distance starts shrinking due to gluinos hitting the sides of the cylinder. 

Figure~\ref{fig: water vesc vs mass} shows the escape-velocity window as a function of gluino mass \(\mg\) at fixed pseudorapidity $\eta=0$. For \(\mg=2~\text{TeV}\) the window spans \((v_{\text{esc}}^{\min},v_{\text{esc}}^{\max})=(0.12,\,0.24)\); for \(\mg=3~\text{TeV}\) it shifts downward to \((0.11,\,0.21)\). The \(\sim10\%\) downward shift of both edges follows from \eq{\ref{eq: stopping length vs mass}}: holding the stopping length \(\xstop\) fixed and substituting the nonrelativistic relation \(K_0=\tfrac{1}{2}\mg\,v_{\text{esc}}^2\), we can solve for the escape velocity,
\begin{align}
v_{\text{esc}}=\left(\frac{8\,\xi C\kappa\,\xstop}{\mg}\right)^{1/4}~.
\end{align}
Thus \(v_{\text{esc}}(2~\text{TeV})/v_{\text{esc}}(3~\text{TeV})=(3/2)^{1/4}\simeq1.1\), reproducing the numerical results.

\section{Chemical bond limit}
\label{sec:chemical bond}
An extremely heavy gluino would not stop in matter as its weight would break through the chemical bonds of materials beneath it, causing it to fall to the center of Earth. This occurs when the gravitational potential energy $\mg gh$ becomes comparable to the typical chemical bond energy $E_B \sim e^2/4\pi r \sim 10$ eV, where $r \sim 10^{-10}$ m is the typical atomic size, and the relevant height is also the atomic size, $h=r$. We find
\begin{align}
    m = \frac{E_B}{gr} \approx 10^{18}~\text{GeV} ~,
\end{align}
which is far larger than the scale relevant here.

A more stringent limit comes from the maximum centrifugal force beyond which a gluino would break the chemical bonds that hold up the wall of the container in the centrifuge. This can be found by using the effective gravitational acceleration in the above equation $g \to g_{\text{eff}} = 10^5 g$, which gives
\begin{align}
    m = \frac{E_B}{g_{\text{eff}} r} \approx 10^{13}~\text{GeV} \left( \frac{10^5 g}{g_{\text{eff}}}\right) ~.
\end{align}
In addition to the force due to the centrifugal acceleration, another limit comes from the wall absorbing the momentum of the gluino. For the gluino to break a chemical bond, it must exert a pressure $P$ exceeding the bond energy density $\rho = E_B/r^3$. Since the pressure is the force exerted by the gluino per atomic cross sectional area, $P = F/r^2$, we have
\begin{align}
    \frac{F}{r^2} = \frac{E_B}{r^3} ~.
\end{align}
The force $F$ can be estimated by considering the change in momentum of the gluino over a characteristic stopping time $t$: $F = \Delta p/ t= \mg v/t$. This gives
\begin{align}
   \mg v \frac{r}{t}  = E_B ~.
\end{align}
Assuming the gluino is stopped over a distance of order an angstrom, we take $v = r / t$, which yields
\begin{align}
    \mg v^2 = E_B ~,
\end{align}
which is essentially a comparison of kinetic energy with the bond energy.

The kinetic energy associated with thermal velocity is independent of the gluino mass. At room temperature, it is given by $\frac{1}{2} \mg \vth^2 \approx  300~\text{K} = 0.025 ~\text{eV}$,
which of course does not exceed the chemical bond limit. In Appendix~\ref{sec:equilibriumTime}, we find that the drift velocity $v_{\text{drift}}$ [\eq{\ref{eq: vdrift}}] does depend on the gluino mass, but for $\mg = 3~\text{TeV}$, it remains much smaller than the thermal velocity. The critical mass at which the drift velocity breaks chemical bond is found by solving $\mg v_{\text{drift}}^2 = E_B $, which gives
\begin{align}
    \mg \sim 10^6~\text{GeV} \left( \frac{g_{\text{eff}}}{10^5 g} \right)^{-2} ~.
\end{align}

\section{Equilibration time}
\label{sec:equilibriumTime}
We aim to determine the timescale over which the gluino reaches its equilibrium distribution within the centrifuge. This equilibration time is controlled by the faster of two processes: thermal diffusion (a random walk) and drift due to centrifugal acceleration (a biased random walk).

\subsection{Diffusion time}
Neglect the net drift arising from centrifugal acceleration for the moment. We first compute the timescale over which the gluino reaches its equilibrium distribution through diffusion alone. Since the gluino is much heavier than the surrounding atoms, its motion is ballistic, meaning its direction of motion is only randomized over many collisions~\cite{Landau_physical_kinetics}. So we use the diffusion equation~\cite{diffusionNote} formulated in terms of the momentum-transfer mean free path $\lambdatr$, which characterizes the average distance over which the gluino’s direction is effectively randomized.

The root mean square distance traversed by the gluino after a time $t$ is given by $d = \lambdatr \sqrt{t/\tautr}$, where $\tautr$ is the momentum-transfer mean free time. Since the thermal velocity is given by $\vth= \lambdatr/\tautr$, we have $d = \sqrt{\lambdatr \vth t}$. By equipartition theorem, we have $\mg v^2_{\text{th}}/2= 3 T/2$, and the transfer mean free path is related to the number density $n$ and the momentum-transfer cross section $\sigmatr$ via $\lambdatr = (n \sigmatr)^{-1}$. Substituting these relations yields
\begin{align}
    d^2 = \frac{t}{n\sigmatr} \sqrt{\frac{3T}{\mg}} ~.
\end{align}

We model the scattering between the anomalously heavy atom and surrounding normal atoms to be hard-sphere scattering, with a total geometric cross section $\sigma \sim \sigma_{\text{geom}} = \pi (2a)^2$, where $a=10^{-10}$~m is the typical atomic size. It can be shown that while in the center of mass frame the transfer cross section is the same as the geometric cross section (which is frame independent), in the lab frame in which the normal atom is initially stationary, the transfer cross section is suppressed by a factor of their mass ratio:
\begin{align}
    \sigmatr \approx \frac{m_{\text{atom}}}{\mg} \sigma_{\text{tr,CM}} = \frac{m_{\text{atom}}}{\mg} \sigma_{\text{geom}} ~.
\end{align}

To reach equilibrium, the gluino must have sufficient time to diffuse across the entire length of the test tube; that is, it must travel an average distance \( d = \rmax \). For $\mg = 3$ TeV and number density of liquid water $n = 3 \times10^{22}~\text{cm}^{-3}$, this sets a diffusion time of
\begin{align}
   t_{\text{diff}} &= \frac{m_{\text{atom}}}{\mg} n 4 \pi a^2 \rmax^2 \sqrt{\frac{\mg}{3T}} \\
   &= \frac{\rho}{\sqrt{3T\mg}} 4 \pi a^2 \rmax^2  \\
   &= 5700 ~\text{s} \left( \frac{3~\text{TeV}}{\mg}\right)^{1/2} \left( \frac{\rmax}{11~\text{cm}}\right)^2~.
\end{align}
where $\rho$ is the density of water.

\subsection{Drift time}
To account for the biased random walk induced by centrifugal acceleration, we adopt a modified form of the Drude model~\cite{Purcell}. As long as the resulting drift velocity \( \vdrift \) remains much smaller than the thermal velocity, \( \vth \approx 10^{-7} \), we may assume that the transfer mean free time, \( \tautr = \lambdatr/\vth \), remains unchanged from the purely diffusive case. Over each interval \( \tautr \), the gluino experiences an outward acceleration due to the centrifugal force and acquires a drift velocity given by
\begin{align}
    \vdrift &\approx (\omega^2 \rmax) \tautr \\
    &= (\omega^2 \rmax) \frac{1}{n \sigmatr \vth} \\
    &= 3 \times 10^{-12} \left( \frac{\mg}{3~\text{TeV}} \right)^{3/2} \left( \frac{\omega^2 \rmax}{10^5 g} \right) ~, \label{eq: vdrift}
\end{align}
which satisfies our condition $\vdrift \ll \vth$. Note that although the drift velocity is much smaller in magnitude than the thermal velocity, it is consistently directed outward, in contrast to diffusion, which is inherently random in direction.
The characteristic time for a gluino to drift across the test tube is then
\begin{align}
    t_{\text{drift}} = \frac{\rmax}{\vdrift} \approx 100~\text{s}  \left( \frac{3~\text{TeV}}{\mg} \right)^{3/2}~.
\end{align}
This timescale therefore determines the equilibration time \( \teq \) as we are in the drift-dominated regime.

\section{Centrifuge equilibrium distribution}
\label{sec:equilbiriumDistribution}

Let $c(r)$ denote the concentration of heavy particles of mass $m$ within the liquid inside a centrifuge, where $r$ is the radial distance from the axis of rotation. Two mechanisms contribute to the particle current in the liquid: diffusion and convection. The diffussive current density is given by $\mathbf{J}_{\text{diff}}= - D \nabla c$, where $D$ is the diffusion coefficient. The convective current density arises from the drift of particles under the effective force due to centrifugal acceleration and is given by $\mathbf{J}_{\text{conv}} = c \mathbf{u}$, where $\mathbf{u}$ is the terminal velocity. The total current density is $\mathbf{J} = \mathbf{J}_{\text{diff}} + \mathbf{J}_{\text{conv}}$. 

The terminal velocity can be found by balancing the centrifugal force $m a_c \hat{r}$ against the buoyant force $-V \rho_0 a_c \hat{r}$ and the frictional force $-fu \hat{r}$, where $V$ is the volume of the heavy particle, $\rho_0$ is the fluid density, $f$ is the drag coefficient, and $a_c = \omega^2 r$ is the centrifugal acceleration. Solving for $u$ gives
\begin{align}
    u &= \frac{(m-V \rho_0)}{f} \omega^2 r \equiv s \omega^2 r ~,
\end{align}
where $s = (m/f)\left(1 - \rho_0/\rho \right)$ is known as the sedimentation coefficient and $\rho$ denotes the density of the heavy particle. 

Substituting the relevant expressions into the continuity equation, $dc/dt = - \nabla \cdot \mathbf{J}$, yields the full dynamical description of sedimentation in a centrifuge, known as the Lamm equation~\cite{Mazumdar_1999}. However, since we are interested solely in the equilibrium distribution, it suffices to impose that the total current density vanishes everywhere: $\mathbf{J}(r) = 0$. This condition implies that diffusive flux exactly balances convective transport due to the centrifugal force:
\begin{align}
    D\frac{dc}{dr}&= s \omega^2 r c ~.
\end{align}
The solution is given by
\begin{align}
        c(r) = c_0 \exp\left( \frac{s}{D} \frac{\omega^2 r^2}{2} \right) ~,
\end{align}
where $c_0$ is the concentration at $r=0$.

The sedimentation coefficient can be written as $s = \mu m_b$, where the mobility is defined as $\mu \equiv 1/f$ and the buoyant mass is defined as $m_b \equiv m \left(1 - \rho_0/\rho \right) $. Since the density $\rho$ of the heavy particle we consider is much greater than the fluid density $\rho_0$, we have $m_b \approx m$. By the Einstein relation~\cite{Kardar_2007}, the diffusion coefficient is equal to the temperature times the mobility, $D = \mu T$, which implies the prefactor in the exponential is in fact
\begin{align}
    \frac{s}{D} = \frac{m}{T} ~.
\end{align}
The equilibrium distribution is then
\begin{align}
     c(r) = c_0 \exp\left(\frac{m\omega^2 r^2/2}{T} \right) ~.
\end{align}
This reproduces the Maxwell-Boltzmann distribution $\exp\left(- U(r)/T \right)$ for a potential energy $U(r)=-\frac{1}{2}m\omega^2 r^2$.

\section{Mass spectrometry: A case study}
\label{sec: mass spec appendix}
Reference~\cite{HeavyWater2} searched for a heavy hydrogenlike particle with charge $+e$ in an enriched sample of heavy water using mass spectrometry (MS). A duoplasmatron was used to ionize a sample of $0.016$~mL of heavy water, producing an ion current of $I=8~\mu\text{A}$. The ions were then accelerated through an electric potential at $V=130$~kV. A magnetic mass selection then filtered out ion masses outside the range 30--1200~GeV. Next, an attenuating carbon foil screened out\footnote{This step exploits the fact that an ion's charge fluctuates as it passes through material~\cite{charge_state_fluctuation}, resulting in an effective charge that can be approximated as~\cite{effective_charge} $Z_{\text{eff}} = Z \left[1-\exp(-125 v Z^{-2/3})\right]$,
where $v$ is the ion velocity.
Hence, regardless of the charge state upon ionization, high-$Z$ ions lose more energy as they pass through the foil, and the foil thickness was designed to only allow the passage of $Z=1$ ions. Our targets of water, silicon, and argon have relatively low $Z$ and we could adapt this method to their atomic numbers.} ions with atomic number $Z>1$ and broke all molecules into constituent atoms. Finally, a time-of-flight measurement was made on the surviving ions, where no positive signal was found. Many aspects of this experiment could be applied or adapted to our case here and conceivably be improved to the required level. 

Starting from $N_{\text{sample}} = 5 \times 10^{20}$, the experiment reached a sensitivity of $\cmin \approx 2 \times 10^{-18}$ for $m=100$~GeV particles (with sensitivity degrading at higher masses), corresponding to an efficiency of $\epsilon \approx 10^{-3}$. Since this can be decomposed into ionization, transmission, and detection efficiency, $\epsilon = \epsilon_i \times \epsilon_t \times \epsilon_d$, we will describe various methods that could independently improve each of these efficiencies.

The ionization efficiency was $\epsilon_i \approx 2.5 \times 10^{-2}$, though Ref.~\cite{HeavyWater2} emphasized this is not an intrinsic limit and described a system that could recycle unionized gas to improve $\epsilon_i$. More recent work has demonstrated a recycling system for noble gas mass spectrometry that gives rise to a 50-fold increase in $\epsilon_i$~\cite{noble_gas_recycling_2010}. If this method is transferable, it would already yield near-perfect ionization efficiency.

The remaining losses for $m=100$~GeV particles were due to large-angle scattering at the attenuating foil, with a transmission efficiency of $\epsilon_t \sim 4 \times 10^{-2}$. While this could conceivably be improved directly, it might be possible to replace this step: the dual purpose of the attenuating foil to break bonds and reject high-$Z$ background could be separated.

The molecular-bond breaking could occur as a preliminary step or as part of the ionization step. We could exploit the molecular fragmentation property of hard ionization techniques~\cite{Ionization_MassSpec_Guide}; for example, electron ionization (EI) at $70$~eV is commonly used to break organic bonds because the de Broglie wavelength of electrons at this energy is $\lambda \approx 0.1$~nm, matching the length of the typical bonds in organic molecules~\cite{70eV_bond_break}. It appears feasible to adapt this method to break other types of molecular bonds.

The filtering of high-$Z$ ions was needed in that experiment because the magnetically selected mass range overlapped with standard model mass range, resulting in a significant background that need to be further reduced. For our purposes, the target particle is in the 2--3~TeV range, far heavier than any standard model background. 
The magnetic field required to separate the heaviest standard model ion can be estimated from the following formula for cyclotron radius $r_c$:
\begin{align}
    B&=\frac{1}{r_c}\sqrt{\frac{2mV}{q}}\\
&= 0.2~\text{T}~\left( \frac{1~\text{m}}{r_c}\right)^{-1} \left( \frac{m}{250~\text{GeV}}\right)^{1/2} \left( \frac{V}{65~\text{kV}}\right)^{1/2} \left( \frac{1}{q}\right)^{1/2} ~.
\end{align}
We see that a modest magnetic field at laboratory length scale is sufficient to remove even the heaviest naturally occurring atoms, Pu-244~\cite{heaviest_element_Scientific_American}, from a straight path. But any hypothetical TeV-scale particles would be essentially unperturbed by the magnetic field\footnote{Although an R-nuclei with $\mg \gtrsim 2$~TeV with multiply charge $q>1$ would be deflected more than the usual $q=1$ case, it would need $q \gtrsim 10$ for it to be indistinguishable from a heavy standard model ion. This can be controlled~\cite{MCP_detector_2014} and large charge is usually suppressed anyway; e.g., $q=2$ is an order of magnitude less likely than $q=1$ for ionized water at low ionization energy due to autodissociation~\cite{doubly_charged_water}.} and continue in a straight path. 

Therefore, we can, in principle, construct a null experiment in which no standard model background is present after the magnetic mass selection stage. In fact, such a magnetic selection for $m>292$~GeV has been conducted in an AMS experiment, in which several hours of background-free data has been achieved~\cite{AMS_background_free_2012}. 

In Ref.~\cite{HeavyWater2}, three detector plates made of thin carbon foils were used to make the final time-of-flight measurement. Although the detection efficiency $\epsilon_d$ was close to 1 for low mass ions, it deteriorated for larger masses, down to $\sim 0.1$ at $m=1.2$~TeV for each plate. Since the ions passed through three such plates, the total sensitivity was approximately $\epsilon_d \sim 10^{-3}$ at this mass.

Fortunately, modern MS achieves markedly higher ion detection efficiency with microchannel plate (MCP) detectors~\cite{MCP_detector_2014}, originally engineered for molecules at the \(m=1\text{--}300~\mathrm{TeV}\) scale in a biological context, but readily repurposable for our case here. The basic principle is that when an ion hits an MCP detector, it triggers a cascade of secondary electrons. Reference~\cite{MCP_detector_2014} experimentally demonstrated that the number of secondary electrons, also known as the secondary electron yield, has a phenomenological scaling
\begin{align} \label{eq: electron yield}
    \gamma \approx 2 \left( \frac{V}{25~\text{kV}} \right)^{3/2} \left( \frac{m}{1~\text{TeV}}\right)^{-1/2}
\end{align}
for a singly charged ion. The ion is detected if at least one secondary electron is emitted. The MCP detection efficiency is given by a Poisson distribution,
\begin{align} \label{eq: epsilon detection}
    \epsilon_d = 1- e^{-\gamma} ~.
\end{align}
It can be shown from \eq{\ref{eq: electron yield}} and \eq{\ref{eq: epsilon detection}} that using an accelerating voltage of $V=65$~kV would enable detection of $m=3$~TeV ions with $\epsilon_d >99\%$.

Note that the MCP detector is a destructive procedure, which means the ion is lost once it is detected. If the magnetic selection step can eliminate all standard model backgrounds, then we could instead use a nondestructive detector; for example, an inductive charge detector detects charge passage via its induced current~\cite{nondesctructive_charge_detection_2008,MCP_detector_2014}.\footnote{Ref.~\cite{nondesctructive_charge_detection_2008} has a charge threshold of $250e$ due to background, but single charge sensitivity might be possible in a background-free environment in principle.} After producing the signal, the heavy ion can be captured, double-checked, and further analyzed.

\clearpage
\onecolumngrid
\section{ATLAS simulation}
\label{sec: simulated materials}

\begin{figure}[t]
    \centering
\includegraphics[width=0.492\textwidth]{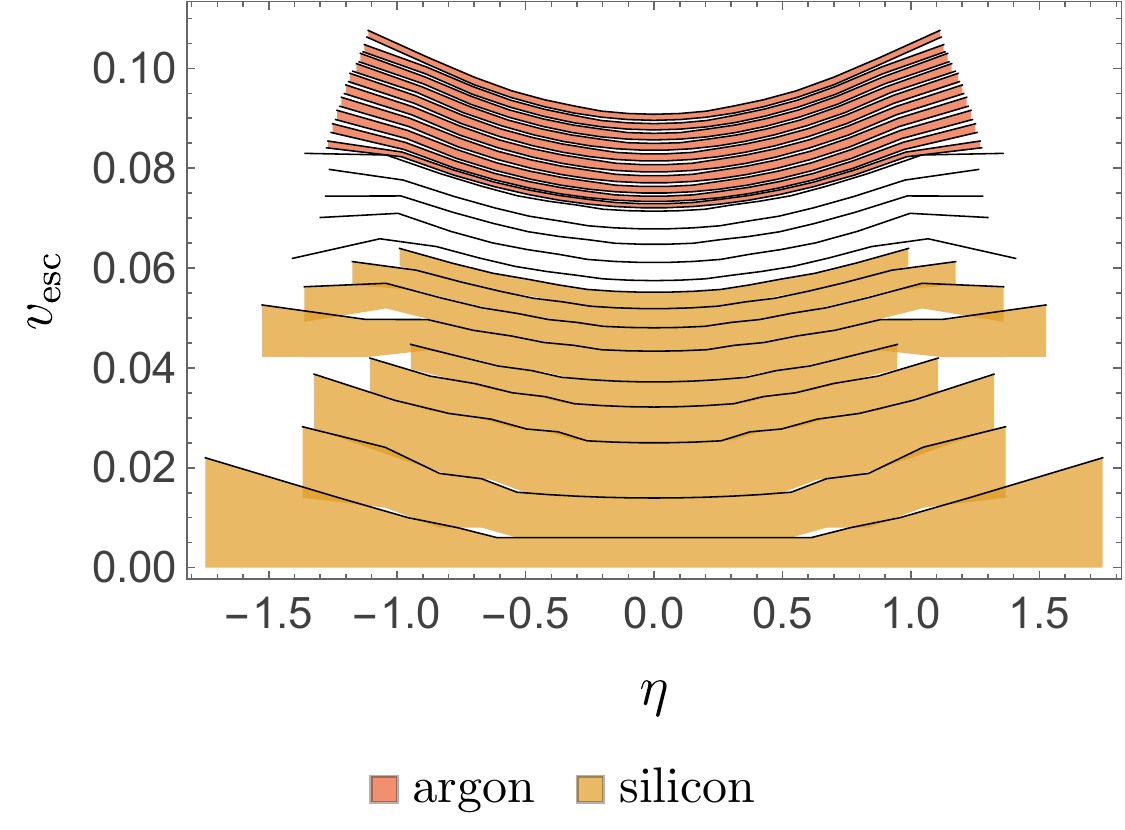}
\caption{Escape velocity \(v_{\rm esc}\) vs.\ pseudorapidity \(\eta\) for 2.5~TeV gluinos in ATLAS barrel for HL-LHC. Here \(v_{\rm esc}\) is the maximum initial velocity for which the gluino still stops within a given layer. Black curves show \(v_{\rm esc}\) for each simulated cylindrical layer. Silicon layers in the ITk and liquid argon layers in the EM calorimeter are highlighted. The liquid argon layers are separated by lead layers.}
\label{fig: vesc vs eta HL}
\end{figure}

\begin{figure}[t]
\centering
\includegraphics[width=0.492\textwidth]{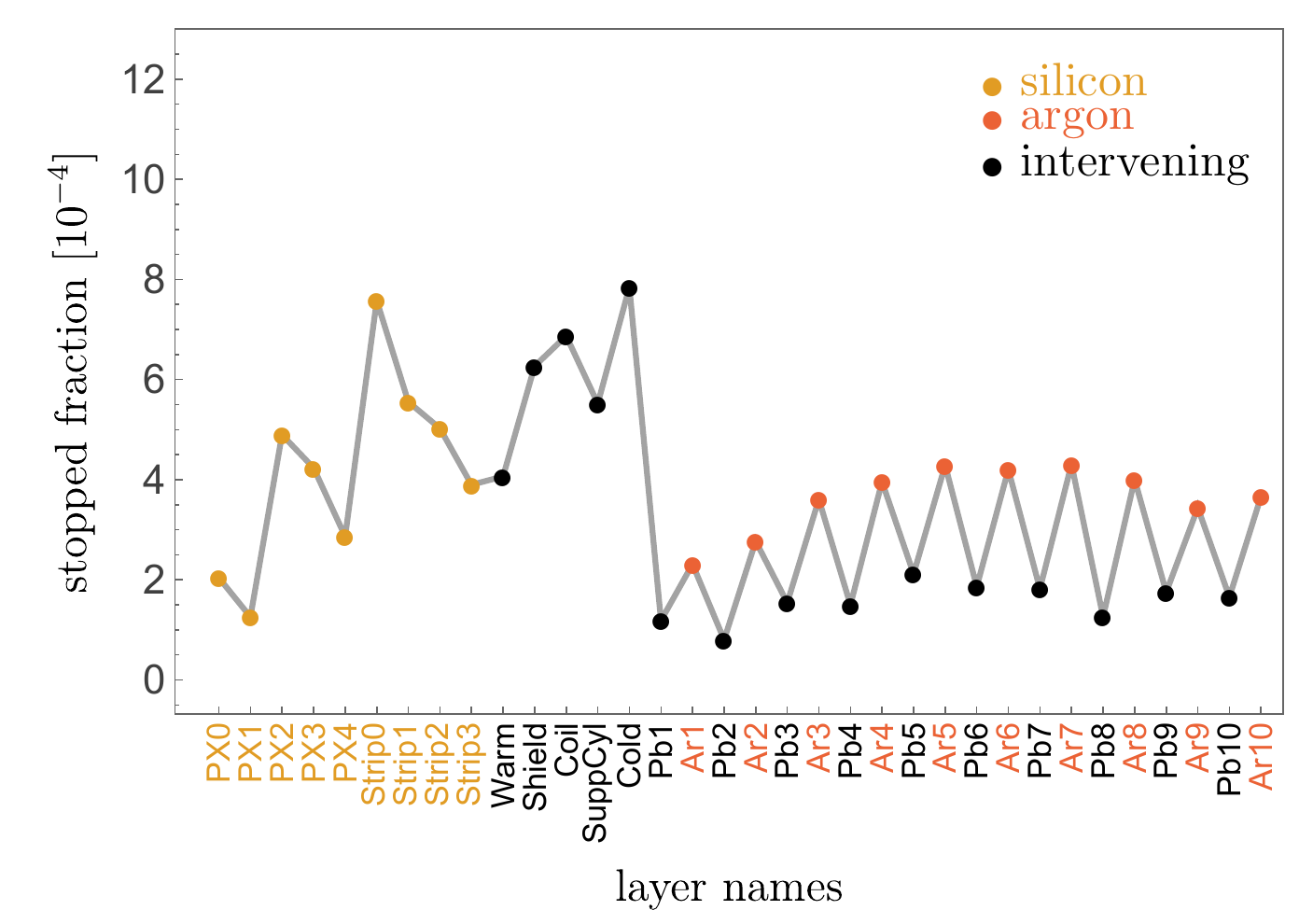}
\caption{Stopped fraction for $2.5~\text{TeV}$ gluinos in each cylindrical layer in the ATLAS barrel in HL-LHC. Silicon layers in the ITk and liquid argon layers in the EM calorimeter are highlighted; all other materials in this region are labeled as ``intervening." Abbreviated layer names are shown on the horizontal axis (Table~\ref{tab:material}). The total stopped fraction of all layers shown is about 1.2\%.
}
\label{fig: stopped fraction ATLAS HLLHC}
\end{figure}

In this appendix, we show some more details of the ATLAS simulation. Figures~\ref{fig: vesc vs eta HL} and \ref{fig: stopped fraction ATLAS HLLHC} show the stopped gluinos results for the HL-LHC, differing from Figs.~\ref{fig: vesc vs eta Run 2} and \ref{fig: stopped fraction ATLAS} because the ID of Run 2/3 is replaced by the ITk system. Table~\ref{tab:material} tabulates the ATLAS materials used in the simulation.

\begin{table*}[h]
  \centering
  \renewcommand{\arraystretch}{1.2}

  \begin{subtable}[t]{0.49\textwidth}
    \centering
    \begin{tabular}{l cc c c c}
      \hline
      System & Name & Mat. & $R$ (mm) & $\Delta Z$ (mm) & $\Delta R$ (mm) \\
      \hline
      \multirow{10}{*}{ID~\cite{ATLAS_ID_Run2}}
        & \textbf{IBL}      & \textbf{Si} &  30 &  400 & 10 \\
        & \textbf{Pixel 1}  & \textbf{Si} &  47 &  400 & 20 \\
        & \textbf{Pixel 2}  & \textbf{Si} &  85 &  400 & 20 \\
        & \textbf{Pixel 3}  & \textbf{Si} & 120 &  400 & 20 \\
        & PSF               & C           & 190 & 1500 & 20 \\
        & \textbf{SCT 1}    & \textbf{Si} & 270 &  750 & 30 \\
        & \textbf{SCT 2}    & \textbf{Si} & 350 &  750 & 30 \\
        & \textbf{SCT 3}    & \textbf{Si} & 420 &  750 & 30 \\
        & \textbf{SCT 4}    & \textbf{Si} & 500 &  750 & 30 \\
        & TRT               & C           & 560 &  712 & 28.6 \\
      \hline
    \end{tabular}
    \subcaption{Run 2/3 Inner Detector (ID).}
    \label{tab:tracker-run2}
  \end{subtable}\hfill
  \begin{subtable}[t]{0.49\textwidth}
    \centering
    \begin{tabular}{l cc c c c}
      \hline
      System & Name & Mat. & $R$ (mm) & $\Delta Z$ (mm) & $\Delta R$ (mm) \\
      \hline
      \multirow{9}{*}{ITk~\cite{ITk_Details_ATLAS_2024}}%
        & \textbf{Pixel 0}  & \textbf{Si} &  34 &  245 & 10 \\
        & \textbf{Pixel 1}  & \textbf{Si} &  99 &  245 & 10 \\
        & \textbf{Pixel 2}  & \textbf{Si} & 160 &  372 & 15 \\
        & \textbf{Pixel 3}  & \textbf{Si} & 228 &  372 & 15 \\
        & \textbf{Pixel 4}  & \textbf{Si} & 291 &  372 & 15 \\
        & \textbf{Strip 0}  & \textbf{Si} & 399 & 1372 & 25 \\
        & \textbf{Strip 1}  & \textbf{Si} & 562 & 1372 & 25 \\
        & \textbf{Strip 2}  & \textbf{Si} & 762 & 1372 & 25 \\
        & \textbf{Strip 3}  & \textbf{Si} & 1000& 1372 & 25 \\
      \hline
    \end{tabular}
    \subcaption{HL-LHC Inner Tracker (ITk).}
    \label{tab:tracker-itk}
  \end{subtable}

  \vspace{0.6em}

  \begin{subtable}[t]{\textwidth}
    \centering
    \begin{tabular}{c cc c c c}
      \hline
      System & Name & Material & $R$ (mm) & $\Delta Z$ (mm) & $\Delta R$ (mm) \\
      \hline
      \multirow{5}{*}{Cryostat~\cite{ATLAS_Solenoid}}
        & Warm Vessel       & Fe & 1152 & 3070 & 13.5 \\
        & Shield            & Al & 1190 & 2650 & 22.5 \\
        & Coil              & Al & 1234 & 2650 & 33 \\
        & Support Cylinder  & Al & 1267 & 2650 & 32.5 \\
        & Cold Vessel       & Al & 1345 & 3267 & 45 \\
      \hline
      \multirow{7}{*}{EM Calorimeter~\cite{ATLAS_Liquid_Argon,ATLAS_EM_Calorimeter_2013}}
        & Lead 1            & Pb          & 1500 & 3200 & 15.7 \\
        & \textbf{Argon 1}  & \textbf{Ar} & 1516 & 3200 & 31.3 \\
        & Lead 2            & Pb          & 1547 & 3200 & 15.7 \\
        & \textbf{Argon 2}  & \textbf{Ar} & 1563 & 3200 & 31.3 \\
        & \vdots            & \vdots      & \vdots& \vdots& \vdots \\
        & Lead 10           & Pb          & 1923 & 3200 & 15.7 \\
        & \textbf{Argon 10} & \textbf{Ar} & 1939 & 3200 & 31.3 \\
      \hline
    \end{tabular}
    \subcaption{Common cryostat and calorimeter for both running periods.}
    \label{tab:common-cryocal}
  \end{subtable}

  \caption{ATLAS barrel material model used in the simulation. Top: comparison between the Run 2/3 inner detector and the HL-LHC inner tracker. Bottom: elements that are identical in both configurations. Silicon and liquid argon are highlighted in bold. Each row represents a cylindrical shell with inner radius $R$, radial thickness $\Delta R$, and axial extent from $-\Delta Z$ to $\Delta Z$ along the beam line. The primary element of the material is also listed.}
  \label{tab:material}
\end{table*}

\twocolumngrid 

\vspace*{\fill}
\clearpage
\bibliography{biblio.bib}

\end{document}